\documentclass{LMCS}

\theoremstyle{plain}

\usepackage{amsmath}
\usepackage{amssymb}
\usepackage{xspace}
\usepackage{latexsym}
\usepackage{verbatim}
\usepackage{graphicx}
\usepackage{hyperref,enumerate}

\def \A             {\mathcal{A}}
\def \aa            {\alpha}

\def \ann           {\ensuremath{\mathsf{ann}}\xspace}

\def \B             {\mathcal{B}}
\def \bottom        {\flat}

\def \coUP          {\textsc{co--UP}\xspace}
\def \CTL           {\emph{CTL}\xspace}
\def \CTLSTAR       {\emph{CTL$^*$}\xspace}
\def \degree        {deg}
\def \edge          {E}

\def \EXP           {{\sc Exptime}\xspace}

\def \EXPTIME       {\textsc{Exptime}\xspace}
\def \F             {\mathcal{F}}
\def \false         {\mathbf{false}}

\def \FSMC          {\textsc{FSMC}\xspace}

\def \GAPT          {\emph{GAPT}\xspace}

\def \Gb            {\Gamma_{\bottom}}

\def \GNPT          {\emph{GNPT}\xspace}
\def \head          {head}
\def \K             {\mathcal{K}}
\def \L             {\mathcal{L}}

\def \M             {\mathcal{M}}
\def \Nat           {\mathbb{N}}
\def \Naturals      {\mbox{I$\!$N}}
\def \NBT           {\emph{NBT}\xspace}
\def \node          {V}
\def \NPT           {\emph{NPT}\xspace}

\def \OPD           {\emph{OPD}\xspace}
\def \P             {\mathcal{P}}

\def \PDMC          {\textsc{PDMC}\xspace}
\def \PDNBT         {\emph{PD--NBT}\xspace}

\def \PDNPT         {\emph{PD--NPT}\xspace}

\def \prom          {\ensuremath{\mathsf{pro}}\xspace}

\def \PTIME         {\textsc{Ptime}\xspace}
\def \rr            {\mathit{root}}
\def \S             {\mathcal{S}}

\def \SPMC          {\textsc{SPMC}\xspace}

\def \strat         {\ensuremath{\mathsf{str}}\xspace}

\def \TAPT          {\emph{$2$APT}\xspace}

\def \TGAPT         {\emph{$2$GAPT}\xspace}

\def \THREEEXPTIME  {\textsc{3Exptime}\xspace}

\def \true          {\mathbf{true}}

\def \TWOEXPTIME    {\textsc{2Exptime}\xspace}

\def \UP            {{\sc UP}}
\def \UP            {\textsc{UP}\xspace}
\def \val           {\mathcal{V}}

\newcommand \tpl[1] {\langle #1 \rangle}
\newcommand \ignore[1] {}

\newtheorem{theorem}{Theorem}
\newtheorem{lemma}{Lemma}

\def\doi{4 (3:1) 2008}
\lmcsheading%
{\doi}
{1--21}
{}
{}
{Sep.~24, 2007}
{Jul.~29, 2008}
{}   

\begin{document}

\title{Enriched $\mu$--Calculi Module Checking\rsuper *}

\author[A.~Ferrante] {Alessandro Ferrante\rsuper a}
\address{{\lsuper{a,c}}Universit\`a di Salerno, Via Ponte don Melillo, 84084 -
Fisciano (SA), Italy} \email{\{ferrante,parente\}@dia.unisa.it}
\thanks{}

\author[A.~Murano]{Aniello Murano\rsuper b}
\address{{\lsuper b}Universit\`{a} di Napoli ``Federico II'', Dipartimento di
Scienze Fisiche, 80126 Napoli, Italy} \email{murano@na.infn.it}
\thanks{}

\author[M.~Parente]{Mimmo Parente\rsuper c}

\titlecomment{{\lsuper *}The paper is based on \cite{FM07} and \cite{FMP07}}

\begin{abstract}
The model checking problem for open systems has been widely studied in the
literature, for both finite--state (\emph{module checking}) and infinite--state
(\emph{pushdown module checking}) systems, with respect to \CTL and \CTLSTAR.
In this paper, we further investigate this problem with respect to the
$\mu$-calculus enriched with nominals and graded modalities (\emph{hybrid
graded $\mu$-calculus}), in both the finite--state and infinite-state settings.
Using an automata-theoretic approach, we show that \emph{hybrid graded
$\mu$-calculus module checking} is solvable in exponential time, while
\emph{hybrid graded $\mu$-calculus pushdown module checking} is solvable in
double-exponential time. These results are also tight since they match the
known lower bounds for \CTL. We also investigate the module checking problem
with respect to the hybrid graded $\mu$-calculus enriched with inverse programs
(\emph{Fully enriched $\mu$-calculus}): by showing a reduction from the tiling
problem, we show its undecidability. We conclude with a short overview of the
model checking problem for the Fully enriched $\mu$-calculus and the fragments
obtained by dropping at least one of the additional constructs.
\end{abstract}

\maketitle

\section{Introduction}\label{sec:Introduction}
\emph{Model-checking} is a formal method, applied in system design, to
automatically verify the ongoing behavior of \emph{reactive systems}
(\cite{CE81,QS81}). In this verification technique the behavior of a system,
formally described by a mathematical model, is checked against a behavioral
constraint, usually specified by a formula in an appropriate temporal logic
(for a survey, see \cite{CGP99}).

In the process of modeling a system, we distinguish between \emph{closed} and
\emph{open} systems~\cite{HP85}. While the behavior of a closed system is
completely determined by the state of the system, the behavior of an open
system depends on the ongoing interaction with its environment~\cite{Hoa85}. In
model checking open systems, introduced and called \emph{module-checking} in
\cite{KVW01}, one should check the system with respect to arbitrary
environments and should take into account uncertainty regarding the
environment. In such a framework, the open finite--state system is described by
a labeled state--transition graph, called in fact \emph{module}, whose set of
states is partitioned into \emph{system states} (where the system makes a
transition) and \emph{environment states} (where the environment makes a
transition). Given a module $\M$, describing the system to be verified, and a
temporal logic formula $\varphi$, specifying the desired behavior of the
system, module checking asks whether for all possible environments, $\M$
satisfies $\varphi$. Therefore, in module checking it is not sufficient to
check whether the full computation tree obtained by unwinding $\M$ (that
corresponds to the interaction of $\M$ with a maximal environment) satisfies
$\varphi$, but it is also necessary to verify that all trees obtained from the
full computation tree by pruning some subtrees rooted in nodes corresponding to
choices disabled by the environment (those trees represent the interactions of
$\M$ with all the possible environments), satisfy $\varphi$. We collect all
such trees in a set named $exec(\M)$. It is worth noticing that each tree in
$exec(\M)$ represents a ``memoryful'' behavior of the environment. Indeed, the
unwinding of a module $\M$ induces duplication of nodes, which allow different
pruning of subtrees. To see an example, consider a two-drink dispenser machine
that serves, upon customer request, tea or coffee. The machine is an open
system and an environment for the system is an infinite line of thirsty people.
Since each person in the line can prefer both tea and coffee, or only tea, or
only coffee, each person suggests a different disabling of the external
choices. Accordingly, there are many different possible environments to
consider. In \cite{KV97,KVW01}, it has been shown that while for linear--time
logics model and module checking coincide, module checking for specification
given in \CTL and \CTLSTAR is exponentially harder than model checking in the
size of the formula and preserves the linearity in the size of the model.
Indeed, \CTL and \CTLSTAR module checking is \EXPTIME--complete and
\TWOEXPTIME--complete, respectively.

In \cite{BMP05,AMV07}, the module checking technique has been extended to
infinite-state systems by considering \emph{open pushdown systems} (\OPD, for
short). These are pushdown systems augmented with finite information that
allows us to partition the set of configurations into \emph{system} and
\emph{environment} configurations. To see an example of an open pushdown
system, consider the above two-drink dispenser machine, with the additional
constraint that a coffee can be served only if the number of coffees served up
to that time is smaller than that of teas served. Such a machine can be clearly
modeled as an open pushdown system (the stack is used to guarantee the
inequality between served coffees and teas). In \cite{BMP05}, it has been shown
that pushdown module checking is \TWOEXPTIME--complete for \CTL and
\THREEEXPTIME--complete for \CTLSTAR. Thus, for pushdown systems, and for
specification given in \CTL and \CTLSTAR, module checking is exponentially
harder than model checking with respect to the size of the formula, while it
preserves the exponential complexity with respect to the size of the
model~\cite{Wal96,Wal00}.

Among the various formalisms used for specifying properties, a valid candidate
is the \emph{$\mu$-calculus}, a very powerful propositional modal logic
augmented with least and greatest fixpoint operators \cite{Koz83} (for a recent
survey, see also \cite{BS06}). The \emph{Fully enriched
$\mu$--calculus}~\cite{BP04} is the extension of the $\mu$--calculus with
\emph{inverse programs}, \emph{graded modalities}, and \emph{nominals}.
Intuitively, inverse programs allow us to travel backwards along accessibility
relations \cite{Var98}, nominals are propositional variables interpreted as
singleton sets \cite{SV01}, and graded modalities enable statements about the
number of successors of a state \cite{KSV02}. By dropping at least one of the
additional constructs, we get a \emph{fragment} of the Fully enriched
$\mu$-calculus. In particular, by inhibiting backward modalities we get the
fragment we call \emph{hybrid graded $\mu$-calculus}. In~\cite{BP04}, it has
been shown that satisfiability is undecidable in the \emph{Fully enriched
$\mu$--calculus}. On the other hand, it has been shown in \cite{SV01,BLMV06}
that satisfiability for each of its fragments is decidable and
\EXPTIME-complete (for more details, see also \cite{BLMV08}). The upper bound
result is based on an automata--theoretic approach via \emph{two-way graded
alternating parity tree automata (\TGAPT)}, along with the fact that each
fragment of the Fully enriched $\mu$-calculus enjoys the \emph{quasi-forest
model property}. Intuitively, \TGAPT generalize alternating automata on
infinite trees as inverse programs and graded modalities enrich the standard
$\mu$--calculus: \TGAPT can move up to a node's predecessor and move down to
\emph{at least $n$} or \emph{all but $n$} successors. Moreover, a quasi-forest
is a forest where nodes can have roots as successors and having quasi-forest
model property means that any satisfiable formula has a quasi-forest as model.
Using \TGAPT and the quasi-forest model property, it has been shown in
\cite{SV01,BLMV06} that given a formula $\varphi$ of a fragment of the Fully
enriched $\mu$-calculus, it is possible to construct a \TGAPT accepting all
trees encodings\footnote{Encoding is done by using a new root node that
connects all roots of the quasi-forest and new atomic propositions which are
used to encode programs and successor nodes corresponding to nominals.}
quasi-forests modeling $\varphi$. Then, the exponential-upper bound follows
from the fact that the emptiness problem for \TGAPT is solvable in \PTIME
\cite{KPV02}.

In this paper, we further investigate the module checking problem and its
infinite-state extension, with respect to the \emph{hybrid graded
$\mu$-calculus}. To see an example of module checking a finite-state open
system w.r.t. an hybrid graded $\mu$-calculus specification, consider again the
above two-drink dispenser machine with the following extra feature: whenever a
customer can choose a drink, he can also call the customer service or the
security service. Suppose also that by taking one of these two new choices, the
drink-dispenser machine stops dispensing drinks, up to the moment the customer
finishes operating with the service. Assume that, for the labeled
state--transition graph modeling the system, we label by $\mathit{choose}$ the
choosing state and by the nominals $o_{c}$ and $o_{s}$ the states in which the
interaction with the customer and the security services start, respectively.
Moreover, suppose we want to check the following property: ``whenever the
customer comes at a choice, he can choose for both the customer and the
security services''. This property can be formalized by the hybrid graded
$\mu$--calculus formula $\nu x. ((choose \rightarrow \tpl{1,call} \ (o_c \vee
o_s)) \wedge [0,-]x)$, which reads ``it is always true that whenever the
drink--dispenser is in the \emph{choose} state, there are at least $2$
\emph{call}--successors in which $(o_c \vee o_s)$ holds''. Clearly, the
considered open system does not satisfy this formula. Indeed, it is not
satisfied by the particular behavior that chooses always the same service.

By exploiting an automata--theoretic approach via tree automata, we show that
hybrid graded $\mu$--calculus module checking is decidable and solvable in
\EXPTIME in the size of the formula and \PTIME in the size of the system. Thus,
as in general, we pay an exponential--time blowup with respect to the model
checking problem (and only w.r.t. the size of the formula) for the module
checking investigation. In particular, we reduce the addressed module checking
problem to the emptiness problem for graded alternating parity tree automata
(\GAPT). In more details, given a model $\M$ and an hybrid graded
$\mu$-calculus formula $\varphi$, we first construct in polynomial time a
B\"uchi tree automaton (\NBT) $\A_{\M}$ accepting $exec(M)$. The construction
of $\A_{\M}$ we propose here extends that used in \cite{KVW01} by also taking
into account that $\M$ must be unwound in a quasi-forest, rather than a tree,
with both nodes and edges labeled. Thus, the set $exec(\M)$ is a set of
quasi-forests, and the automaton $\A_{\M}$ we construct will accept all trees
encodings of all quasi-forests of $exec(\M)$. From the formula side,
accordingly to \cite{BLMV06}, we can construct in a polynomial time a \GAPT
$\A_{\not \models \varphi}$ accepting all models that do not satisfy $\varphi$,
with the intent to check that none of these models are in $exec(\M)$. Thus, we
check that $\M$ models $\varphi$ for every possible choice of the environment
by checking whether the $\L(\A_{\M}) \cap \L(\A_{\not \models \varphi})$ is
empty. The results follow from the fact that an \NBT is a particular case of
\GAPT, which are closed under intersection and have the emptiness problem
solvable in \EXP \cite{BLMV06}. We also show a lower bound matching the
obtained upper bound by using a reduction from the module checking for \CTL,
known to be \EXP-hard.

By exploiting again an automata-theoretic approach, we show that hybrid graded
$\mu$-calculus pushdown module checking is decidable and solvable in
\TWOEXPTIME in the size of the formula and \EXPTIME in the size of the system.
Thus, as in general, with respect to the finite--state model checking case we
pay an exponential--time blowup in the size of both the system and the formula
for the use of pushdown systems, and an another exponential--time blowup in the
size of the formula for the module checking investigation. Our approach allow
us do not take the trivial \TWOEXPTIME result on both the size of the system
and the formula, which can be easily obtained by combining the algorithms
existing in the literature along with that one we introduce in this paper for
the finite--state case. We solve the hybrid graded $\mu$-calculus pushdown
module checking by using a reduction to the emptiness problem for
nondeterministic pushdown parity tree automata (\PDNPT). The algorithm we
propose extends that given for the finite-state case. In particular, given an
\OPD $\S$, a module $\M$ induced by the configurations of $\S$, and an hybrid
graded $\mu$-calculus formula $\varphi$, we first construct in polynomial time
a pushdown B\"uchi tree automaton (\PDNBT) $\A_{\M}$, accepting $exec(\M)$.
From the formula side, accordingly to \cite{BLMV06}, we can construct in a
polynomial time a \GAPT $\A_{\not \models \varphi}$ accepting all models that
do not satisfy $\varphi$. Thus, we can check that $\M$ models $\varphi$ for
every possible choice of the environment by checking whether $\L(\A_{\M}) \cap
\L(\A_{\not \models \varphi})$ is empty. By showing a non-trivial exponential
reduction of \TGAPT into \NPT, we show a \TWOEXPTIME upper bound for the
addressed problem. Since the pushdown module checking problem for \CTL is
\TWOEXPTIME-hard, we get that the addressed problem is then
\TWOEXPTIME-complete.

As regarding the Fully enriched $\mu$-calculus, we also investigate the module
checking problem in a ``rewind'' framework in the following sense. As far as
backward modalities concern, everytime the system goes back to an environment's
node, he is always able to redefine a new pruning choice. Given a module $\M$
and a Fully enriched $\mu$-calculus formula $\varphi$, we solve the rewind
module checking problem by checking that all trees in $exec(\M)$, always taking
the same choice in duplicate environment nodes, satisfy $\varphi$. By showing a
reduction from the tiling problem \cite{Ber66}, we show that the addressed
problem is undecidable.

We conclude the paper with short considerations on the model
checking on all of the fragments of the Fully enriched
$\mu$-calculus. In particular we show the problem to be
\EXPTIME-complete for a pushdown system which is allowed to push
one symbol per time onto the stack, with respect to any fragment
not including the graded modality: for the fragments with the
graded modality, we show a \TWOEXPTIME upper bound.

\ignore{ Ricordarsi di dire qualcosa sul model checking per il
$\mu$-calculus classico citando i lavori di Igor Walukiewicz e
altri (vedi anche citazioni di Igor Walukiewicz )

Ricordarsi di citare la survey sul $\mu$-calculus del 2007 di
Stirling e Bradfield. }

The rest of the paper is organized as follows. In
Section~\ref{sec:Preliminaries}, we give all the necessary preliminaries,
Section~\ref{sec:HybridGradedModuleChecking} contains the definition of module
checking w.r.t. hybrid graded $\mu$-calculus, and
Section~\ref{sec:TreeAutomata} contains definitions and known results about
\TGAPT and \PDNPT. In
Sections~\ref{sec:DecidingHybridGradedModuleChecking}~and~\ref{sec:DecidingHybridGradedPushdownModuleChecking},
we give our main results on module checking for the hybrid graded
$\mu$-calculus. In Section~\ref{sec:FullyEnrichedModuleChecking}, we show the
undecidability result for the Fully enriched module checking and conclude in
Section~\ref{sec:ModelChecking} with some complexity considerations on model
checking with all the fragments of the Fully enriched $\mu$-calculus.

\section{Preliminaries}\label{sec:Preliminaries}
In this section, we recall definitions of labeled forests and hybrid graded
$\mu$--calculus. We refer to \cite{BLMV06} for more technical definitions and
motivating examples.

\subsection{Labeled Forests.}\label{sub:LabeledForests} For a
finite set $X$, we denote the \emph{size} of $X$ by $|X|$, the set
of words over $X$ by $X^*$, the empty word by $\varepsilon$, and
with $X^+$ we denote $X^* \setminus \{\varepsilon\}$. Given a word
$w$ in $X^*$ and a symbol $a$ of $X$, we use $w \cdot a$ to denote
the word $wa$. Let $\Naturals$ be the set of positive integers.
For $n \in \Naturals$, let $\Nat$ denote the set $\{1,2,\ldots,
n\}$. A \emph{forest} is a set $F \subseteq \Nat^+$ such that if
$x \cdot c \in F$, where $x \in \Nat^+$ and $c \in \Nat$, then
also $x \in F$. The elements of $F$ are called \emph{nodes}, and
words consisting of a single natural number are \emph{roots} of
$F$. For each root $r\in F$, the set $T=\{r\cdot x \mid x\in\Nat^*
\mbox{ and } r\cdot x\in F \}$ is a \emph{tree} of $F$ (the tree
\emph{rooted at $r$}). For $x \in F$, the nodes $x \cdot c \in F$
where $c \in \Nat$ are the \emph{successors} of $x$, denoted
$sc(x)$, and $x$ is their \emph{predecessor}. The number of
successors of a node $x$ is called the \emph{degree} of $x$
($deg(x)$). The degree $h$ of a forest $F$ is the maximum of the
degrees of all nodes in $F$ and the number of roots. A forest with
degree $h$ is an \emph{$h$-ary} forest. A full $h$-ary forest is a
forest having $h$ roots and all nodes with degree $h$.

Let $F \subseteq \Nat^+$ be a forest, $x$ a node in $F$, and $c
\in \Nat$. As a convention, we take $x \cdot \varepsilon =
\varepsilon \cdot x = x$, $(x \cdot c) \cdot -1 = x$, and $c \cdot
-1$ as undefined. We call $x$ a \emph{leaf} if it has no
successors. A \emph{path} $\pi$ in $F$ is a word $\pi=x_1 x_2
\ldots$ of $F$ such that $x_1$ is a root of $F$ and for every $x_i
\in \pi$, either $x_i$ is a leaf (i.e., $\pi$ ends in $x_i$) or
$x_{i}$ is a predecessor of $x_{i+1}$. Given two alphabets
$\Sigma_1$ and $\Sigma_2$, a ($\Sigma_1,\Sigma_2$)--labeled forest
is a triple $\tpl{F, \node, \edge}$, where $F$ is a forest, $\node
: F \rightarrow \Sigma_1$ maps each node of $F$ to a letter in
$\Sigma_1$, and $\edge: F \times F \rightarrow \Sigma_2$ is a
partial function that maps each pair $(x, y)$, with $y \in sc(x)$,
to a letter in $\Sigma_2$. As a particular case, we consider a
forest without labels on edges as a $\Sigma_1$--labeled forest
$\tpl{F, \node}$, and a \emph{tree} as a forest containing exactly
one tree. A \emph{quasi--forest} is a forest where each node may
also have roots as successors. For a node $x$ of a quasi--forest,
we set $children(x)$ as $sc(x)\setminus \Nat$. All the other
definitions regarding forests easily extend to quasi--forests.
Notice that in a quasi--forest, since each node can have a root as
successor, a root can also have several predecessors, while every
other node has just one. Clearly, a quasi--forest can always be
transformed into a forest by removing root successors.

\subsection{Hybrid Graded $\mu$--Calculus.}\label{sub:FullyEnrichedMuCalculus}
Let $\mathit{AP}$, $\mathit{Var}$, $\mathit{Prog}$, and $\mathit{Nom}$ be
finite and pairwise disjoint sets of \emph{atomic propositions},
\emph{propositional variables}, \emph{atomic programs} (which allow to travel
the system along accessibility relations), and \emph{nominals} (which are
particular atomic propositions interpreted as singleton sets). The set of
\emph{hybrid graded $\mu$--calculus} formulas is the smallest set such that
\begin{enumerate}[$\bullet$]
\item
\textbf{true} and \textbf{false} are formulas;

\item
$p$ and $\neg p$, for $p \in AP$, are formulas;

\item
$o$ and $\neg o$, for $o \in \mathit{Nom}$, are formulas;

\item
$x \in \mathit{Var}$ is a formula;

\item
if $\varphi_1$ and $\varphi_2$ are formulas, $\alpha \in Prog$, $n$ is a non
negative integer, and $y \in Var$, then the following are also formulas:
    $$\varphi_1 \vee \varphi_2, \varphi_1 \wedge \varphi_2, \tpl{n,\alpha} \varphi_1,
            [n, \alpha] \varphi_1,~ \mu y. \varphi_1(y)\text{, and } \nu y. \varphi_1(y).$$
\end{enumerate}

Observe that we use positive normal form, i.e., negation is applied only to
atomic propositions.

We call $\mu$ and $\nu$ \emph{fixpoint operators}. A propositional variable $y$
occurs \emph{free} in a formula if it is not in the scope of a fixpoint
operator. A \emph{sentence} is a formula that contains no free variables. We
refer often to the \emph{graded modalities} $\tpl{n, \alpha} \varphi_1$ and
$[n, \alpha] \varphi_1$ as respectively \emph{atleast formulas} and
\emph{allbut formulas} and assume that the integers in these operators are
given in binary coding: the contribution of $n$ to the length of the formulas
$\tpl{n,\alpha}\varphi$ and $[n,\alpha]\varphi$ is $\lceil \log n \rceil$
rather than~$n$.

The semantics of the hybrid graded $\mu$--calculus is defined with respect to a
\emph{Kripke structure}, i.e., a tuple $\K = \tpl{W,W_0,R,L}$ where $W$ is a
non--empty set of \emph{states}, $W_0 \subseteq W$ is the set of initial
states, $R : \mathit{Prog} \rightarrow 2^{W \times W}$ is a function that
assigns to each atomic program a transition relation over $W$, and $L:AP \cup
\mathit{Nom} \rightarrow 2^{W}$ is a labeling function that assigns to each
atomic proposition and nominal a set of states such that the sets assigned to
nominals are singletons and subsets of $W_0$. If $(w,w')\in R(\alpha)$, we say
that $w'$ is an \emph{$\alpha$--successor} of $w$. Informally, an
\emph{atleast} formula $\tpl{n,\alpha}\varphi$ holds at a state $w$ of $\K$ if
$\varphi$ holds in at least $n+1$ $\alpha$--successors of $w$. Dually, the
\emph{allbut} formula $[n, \alpha] \varphi$ holds in a state $w$ of $\K$ if
$\varphi$ holds in all but at most $n$ $\alpha$--successors of $w$. Note that
$\neg \tpl{n,\alpha} \varphi$ is equivalent to $[n, \alpha]\neg \varphi$, and
the modalities $\langle \alpha \rangle \varphi$ and $[\alpha] \varphi$ of the
standard $\mu$--calculus can be expressed as $\tpl{0,\alpha} \varphi$ and
$[0,\alpha] \varphi$, respectively.

To formalize semantics, we introduce valuations. Given a Kripke
structure $\K = \langle W$, $W_0$, $R$, $L \rangle$ and a set
$\{y_1,\ldots, y_n \}$ of variables in $\mathit{Var}$, a
\emph{valuation} $\val: \{y_1,\ldots,y_n \} \rightarrow 2^{W}$ is
an assignment of subsets of $W$ to the variables $y_1,\ldots,
y_n$.  For a valuation $\val$, a variable $y$, and a set $W'
\subseteq W$, we denote by $\val[y \leftarrow W']$ the valuation
obtained from $\val$ by assigning $W'$ to $y$. A formula $\varphi$
with free variables among $y_1, \ldots, y_n$ is interpreted over
$\K$ as a mapping $\varphi^{\K}$ from valuations to $2^{W}$, i.e.,
$\varphi^{\K}(\val)$ denotes the set of points that satisfy
$\varphi$ under valuation $\val$. The mapping $\varphi^{\K}$ is
defined inductively as follows:
\begin{enumerate}[$\bullet$]
\item
$\textbf{true}^{\K}(\val) = W$ and $\textbf{false}^{\K}(\val) =
\emptyset $;

\item
for $p \in AP \cup \mathit{Nom}$, we have $p^{\K}(\val) = L(p)$
and $(\neg p)^{\K}(\val) = W \setminus L(p)$;

\item
for $y \in \mathit{Var}$, we have $y^{\K}(\val) = \val(y)$;

\item
$(\varphi_1 \wedge \varphi_2)^{\K}(\val) = \varphi_1^{\K}(\val)
\cap \varphi_2^{\K}(\val)$ and $(\varphi_1 \vee
\varphi_2)^{\K}(\val) = \varphi_1^{\K}(\val) \cup
\varphi_2^{\K}(\val)$;

\item
$(\tpl{n,\alpha} \varphi)^{\K}(\val)= \{w : |\{w'\in W : (w,w')\in
R(\alpha) \text{ and } w' \in \varphi^{\K}(\val)\}|\geq n+1 \}$;

\item
$([n,\alpha] \varphi)^{\K}(\val)= \{w : |\{w'\in W : (w,w')\in
R(\alpha) \text{ and } w' \not \in \varphi^{\K}(\val)\}|\leq n
\}$;

\item
$(\mu y.\varphi(y))^{k}(\val)= \bigcap \{W'\subseteq W :
\varphi^{\K}([y \leftarrow W']) \subseteq W'\}$;

\item
$(\nu y.\varphi(y))^{k}(\val)= \bigcup \{W'\subseteq W : W'
\subseteq \varphi^{\K}([y \leftarrow W'])\}$.
\end{enumerate}

For a state $w$ of a Kripke structure $\K$, we say that $\K$ \emph{satisfies}
$\varphi$ at $w$ if $w \in \varphi^{\K}$. In what follows, a formula $\varphi$
\emph{counts} up to $b$ if the maximal integer in \emph{atleast} and
\emph{allbut} formulas used in $\varphi$ is $b-1$.

\section{Hybrid graded $\mu$-calculus module Checking}\label{sec:HybridGradedModuleChecking}
In this paper we consider open systems, i.e., systems that
interact with their environment and whose behavior depends on this
interaction. The (global) behavior of such a system is described
by a \emph{module} $\M = \tpl{W_s,W_e,W_0,R,L}$, which is a Kripke
structure where the set of states $W = W_s \cup W_e$ is
partitioned in \emph{system states} $W_s$ and \emph{environment
states} $W_e$.

Given a module $\M$, we assume that its states are ordered and the
number of successors of each state $w$ is finite. For each $w \in
W$, we denote by $succ(w)$ the ordered tuple (possibly empty) of
$w$'s $\alpha$-successors, for all $\alpha \in Prog$. When $\M$ is
in a system state $w_s$, then all states in $succ(w_s)$ are
possible next states. On the other hand, when $\M$ is in an
environment state $w_e$, the possible next states (that are in
$succ(w_e)$) depend on the current environment. Since the behavior
of the environment is not predictable, we have to consider all the
possible sub--tuples of $succ(w_e)$. The only constraint, since we
consider environments that cannot block the system, is that not
all the transitions from $w_e$ are disabled.

The set of all (maximal) computations of $\M$, starting from
$W_0$, is described by a $(W,Prog)$--labeled quasi--forest
$\tpl{F_{\M},\node_{\M},\edge_{\M}}$, called \emph{computation
quasi--forest}, which is obtained by unwinding $\M$ in the usual
way. The problem of deciding, for a given branching--time formula
$\varphi$ over $AP\cup Nom$, whether $\tpl{F_{\M}, L \circ
\node_{\M},\edge_{\M}}$ satisfies $\varphi$ at a root node,
denoted $\M \models \varphi$, is the usual \emph{model--checking
problem} \cite{CE81,QS81}. On the other hand, for an open system
$\M$, the quasi--forest $\tpl{F_{\M}, \node_{\M},\edge_{\M}}$
corresponds to a very specific environment, i.e., a maximal
environment that never restricts the set of its next states.
Therefore, when we examine a branching--time formula $\varphi$
w.r.t. $\M$, the formula $\varphi$ should hold not only in
$\tpl{F_{\M},\node_{\M},\edge_{\M}}$, but in all quasi-forests
obtained by pruning from $\tpl{F_{\M},\node_{\M},\edge_{\M}}$
subtrees rooted at children of environment nodes, as well as
inhibiting some of their jumps to roots (that is, successor nodes
labeled with nominals), if there are any. The set of these
quasi--forests, which collects all possible behaviors of the
environment, is denoted by $exec(\M)$ and is formally defined as
follows. A quasi--forest $\tpl{F,\node,\edge} \in exec(\M)$ iff

\begin{enumerate}[$\bullet$]
\item
for each $w_i \in W_0$, we have $\node(i)=w_i$;

\item
for each $x \in F$, with $\node(x) = w$, $succ(w) = \tpl{w_1,
\ldots, w_n, w_{n+1}, \ldots, w_{n+m}}$, and $succ(w) \cap W_0 =
\tpl{w_{n+1}, \ldots, w_{n+m}}$, there exists $S = \tpl{w'_1,
\ldots, w'_p, w'_{p+1}, \ldots, w'_{p+q}}$ sub-tuple of $succ(w)$
such that $p+q \geq 1$ and the following hold:

\begin{enumerate}[$-$]
\item
$S = succ(w)$ if $w \in W_s$;

\item
$children(x) = \{x \cdot 1, \ldots, x \cdot p\}$ and, for $1 \leq
j \leq p$, we have $\node(x \cdot j) = w'_j$ and $\edge(x, x \cdot
j) = \alpha$ if $(w, w'_j) \in R(\alpha)$;

\item
for $1 \leq j \leq q$, let $x_j \in \Nat$ such that $\node(x_j) =
w'_{p+j}$, then $\edge(x, x_j) = \alpha$ if $(w, w'_{p+j}) \in
R(\alpha)$.
\end{enumerate}
\end{enumerate}
In the following, we consider quasi--forests in $exec(\M)$ as labeled with
$(2^{AP \cup Nom}, \allowbreak Prog)$, i.e., taking the label of a node $x$ as
$L(\node(x))$. For a module $\M$ and a formula $\varphi$ of the hybrid graded
$\mu$--calculus, we say that $\M$ \emph{reactively} satisfies $\varphi$,
denoted $\M\models_r\varphi$ (where ``r'' stands for \emph{reactively}), if all
quasi-forests in $exec(\M)$ satisfy $\varphi$. The problem of deciding whether
$\M \models_r \varphi$ is called \emph{hybrid graded $\mu$--calculus module
checking}.

\subsection{Open Pushdown Systems
(\OPD)}\label{sub:OpenPushdownSystems} An \OPD over
$AP$, $Nom$ and $Prog$
is a tuple $\S = \tpl{Q, \Gamma, \bottom, C_0, \Delta, \rho_1,
\rho_2, Env}$, where $Q$ is a finite set of (control)
\emph{states}, $\Gamma$ is a finite \emph{stack alphabet},
$\bottom \not \in \Gamma$ is the \emph{stack bottom symbol}. We
set $\Gb= \Gamma \cup \{\bottom \}$, $Conf = Q \times (\Gamma^*
\cdot \bottom)$ to be the set of \emph{(pushdown) configurations},
and for each configuration $(q,A \cdot \gamma)$, we set $top((q,A
\cdot \gamma))=(q,A)$ to be a \emph{top configuration}. The
function $\Delta : Prog \rightarrow 2^{(Q \times \Gb) \times (Q
\times \Gb^*)}$ is a finite set of transition rules such that
$\bottom$ is always present at the bottom of the stack and nowhere
else (thus whenever $\bottom$ is read, it is pushed back). Note
that we make this assumption also about the various pushdown
automata we use later. The set $C_0 \subseteq Conf$ is a finite
set of \emph{initial configurations}, $\rho_1: AP \rightarrow 2^{Q
\times \Gb}$ and $\rho_2: Nom \rightarrow C_0$ are labeling
functions associating respectively to each atomic proposition $p$
a set of top configurations in which $p$ holds and to each nominal
exactly one initial configuration. Finally, $Env \subseteq Q
\times \Gb$ specifies the set of \emph{environment
configurations}. The size $|\S|$ of $\S$ is $|Q| + |\Delta| +
|\Gamma|$.

The \OPD moves in accordance with the transition relation
$\Delta$. Thus, $((q,A),\allowbreak (q',\gamma)) \in
\Delta(\alpha)$ implies that if the \OPD is in state $q$ and the
top of the stack is $A$, it can move along with an
\emph{$\alpha$--transition} to state $q'$, and substitute $\gamma$
for $A$. Also note that the possible operations of the system, the
labeling functions, and the designation of configurations as
environment configurations, are all dependent only on the current
control state and the top of the stack.

An \OPD $\S$ induces a module $\M_{\S} = \tpl{W_s, W_e, W_0, R,
L}$, where:
\begin{enumerate}[$\bullet$]
\item
$W_s \cup W_e = Conf$, i.e. the set of pushdown configurations,
and $W_0 = C_0$;

\item
$W_e = \{c \in Conf\ |\ top(c) \in Env\}$.

\item
$((q, A \cdot \gamma),(q', \gamma' \cdot \gamma)) \in R(\alpha)$
\emph{iff} there is $((q, A),(q', \gamma')) \in \Delta(\alpha)$;

\item
$L(p)=\{c \in Conf\ |\ top(c) \in \rho_1(p)\}$ for $p \in AP$;
$L(o)=\rho_2(o)$ for $o \in Nom$.
\end{enumerate}
The \emph{hybrid graded ($\mu$-calculus) pushdown module checking}
problem is to decide, for a given \OPD $\S$ and an enriched
$\mu$--calculus formula $\varphi$, whether $\M_{\S} \models_r
\varphi$.

\section{Tree Automata}\label{sec:TreeAutomata}
\subsection{Two-way Graded Alternating Parity Tree Automata (\TGAPT)}\label{sub:Tgapt}
These automata have been introduced and deeply investigated in \cite{BLMV06}.
In this section we just recall the main definitions and results and refer to
the literature for more details. Intuitively, \TGAPT are an extension of
nondeterministic tree automata in such a way that a \TGAPT can send several
copies of itself to the same successor (\emph{alternating}), send copies of
itself to the predecessor (\emph{two-way}), specify a number $n$ of successors
to which copies of itself are sent (\emph{graded}), and accept trees along with
a \emph{parity acceptance condition}. To give a more formal definition, let us
recall some technicalities from \cite{BLMV06}.

For a given set $Y$, let $B^{+}(Y)$ be the set of positive Boolean formulas
over $Y$ (i.e., Boolean formulas built from elements in $Y$ using $\wedge$ and
$\vee$), where we also allow the formulas \textbf{true} and \textbf{false} and
$\wedge$ has precedence over $\vee$. For a set $X \subseteq Y$ and a formula
$\theta \in B^{+}(Y)$, we say that $X$ satisfies $\theta$ iff assigning
\textbf{true} to elements in $X$ and assigning \textbf{false} to elements in $Y
\setminus X$ makes $\theta$ \textbf{true}.  For $b>0$, let $\tpl{[b]} = \{
\tpl{0}, \tpl{1}, \ldots, \tpl{b}\}$, $[[b]]= \{[0],[1],\ldots,[b]\}$, and $D_b
= \tpl{[b]} \cup [[b]] \cup \{-1, \varepsilon \}$. Intuitively, $D_b$ collects
all possible directions in which the automaton can proceed.

Formally, a \TGAPT on $\Sigma$-labeled trees is a tuple $\A = \langle \Sigma$,
$b$, $Q$, $\delta$, $q_0$, $\F \rangle$, where $\Sigma$ is the input alphabet,
$b > 0$ is a counting bound, $Q$ is a finite set of states, $\delta: Q \times
\Sigma \rightarrow B^{+}(D_b \times Q)$ is a transition function, $q_0 \in Q$
is an initial state, and $\F$ is a parity acceptance condition (see below).
Intuitively, an atom $(\tpl{n}, q)$ (resp.\ $([n], q)$) means that $\A$ sends
copies in state $q$ to $n + 1$ (resp.\ all but $n$) different successors of the
current node, $(\varepsilon, q)$ means that $\A$ sends a copy (in state $q$) to
the current node, and $(-1, q)$ means that $\A$ sends a copy to the predecessor
of the current node. A \emph{run} of $\A$ on an input $\Sigma$-labeled tree
$\tpl{T,V}$ is a tree $\tpl{T_r,r}$ in which each node is labeled by an element
of $T \times Q$. Intuitively, a node in $T_r$ labeled by $(x, q)$ describes a
copy of the automaton in state $q$ that reads the node $x$ of $T$. Runs start
in the initial state and satisfy the transition relation. Thus, a run
$\tpl{T_r,r}$ with root $z$ has to satisfy the following: (\emph{i}) $r(z) =
(1, q_0)$ for the root $1$ of $T$ and (\emph{ii}) for all $y \in T_r$ with
$r(y) = (x, q)$ and $\delta(q, V(x)) = \theta$, there is a (possibly empty) set
$S \subseteq D_b \times Q$, such that $S$ satisfies $\theta$, and for all $(d,
s) \in S$, the following hold:

\begin{enumerate}[$\bullet$]
    \item
    If $d \in \{-1, \varepsilon \} $, then $x \cdot d$ is defined, and there is $j
    \in \Nat$ such that $y \cdot j \in T_r$ and $r(y \cdot j) = (x \cdot d, s)$;

    \item
    If $d = \tpl{n}$, there are at least $n+1$ distinct indexes $i_1, \ldots,
    i_{n+1}$ such that for all $1 \leq j \leq n+1$, there is $j' \in \Nat$ such
    that $y \cdot j' \in T_r$, $x\cdot i_j\in T$, and $r(y \cdot j') = (x \cdot
    i_j, s)$.

    \item
    If $d = [n]$, there are at least $\degree(x)-n$ distinct indexes $i_1, \ldots,
    i_{\degree(x)-n}$ such that for all $1 \leq j \leq \degree(x)-n$, there is $j'
    \in \Nat$ such that $y \cdot j' \in T_r$, $x\cdot i_j\in T$, and $r(y \cdot j')
    = (x \cdot i_j, s)$.
\end{enumerate}

Note that if $\theta = \textbf{true}$, then $y$ does not need to
have successors. This is the reason why $T_r$ may have leaves.
Also, since there exists no set $S$ as required for $\theta =
\textbf{false}$, we cannot have a run that takes a transition with
$\theta = \textbf{false}$.

A run $\tpl{T_r,r}$ is \emph{accepting} if all its infinite paths
satisfy the acceptance condition. In the parity acceptance
condition, $\F$ is a set $\{F_1, \ldots , F_k\}$ such that $F_1
\subseteq \ldots \subseteq F_k = Q$ and $k$ is called the
\emph{index} of the automaton. An infinite path $\pi$ on $T_r$
\emph{satisfies} $\F$ if there is an even $i$ such that $\pi$
contains infinitely many states from $F_i$ and finitely many
states from $F_{i-1}$. An automaton \emph{accepts} a tree iff
there exists an accepting run of the automaton on the tree. We
denote by $\L(\A)$ the set of all $\Sigma$-labeled trees that $\A$
accepts. The \emph{emptiness} problem for an automaton $\P$ is to
decide whether $\L(\P)=\emptyset$.

A \TGAPT is a \GAPT (i.e., ``\emph{one--way}'') if $\delta: Q
\times \Sigma \rightarrow B^{+}(D_b \setminus \{-1\} \times Q)$
and a \TAPT (i.e., ``\emph{non-graded}'')  if $\delta : Q \times
\Sigma \rightarrow B^+(\{-1, \varepsilon, 1, \ldots, h\} \times
Q)$. As a particular case of \TAPT, we also consider
nondeterministic parity tree automata (\NPT) \cite{KVW00}.
Formally, an \NPT on $\Sigma$-labeled trees is a tuple $\A =
\langle \Sigma$, $D$, $Q$, $\delta$, $q_0$, $\F \rangle$, where
$\Sigma, Q, q_0$, and $\F$ are as in \TAPT, $D$ is a finite set of
\emph{branching degree} and $\delta: Q \times \Sigma \times D
\rightarrow 2^{Q^*}$ is a transition function satisfying
$\delta(q, \sigma, d) \subseteq Q^d$, for each $q \in Q$, $\sigma
\in \Sigma$, and $d \in D$. Finally, we also consider
\emph{B\"uchi acceptance condition} $\F \subseteq Q$, which simply
is a special parity condition $\{\emptyset, \F, Q\}$. Thus, we use
in the following the acronym \NBT to denote nondeterministic
B\"uchi tree automata on $\Sigma$-labeled trees.

The following results on \TGAPT will be useful in the rest of the
paper.

\begin{theorem}\cite{BLMV06}\label{the:TgaptEmptiness}
The emptiness problem for a \GAPT $\A = \tpl{\Sigma, b, Q, \delta,
q_0, \F}$ can be solved in time linear in the size of $\Sigma$ and
$b$, and exponential in the index of the automaton and number of
states.
\end{theorem}

\begin{lemma}\cite{BLMV06}\label{lem:TgaptIntersection}
Given two \GAPT $\A_1$ and $\A_2$, there exists a \GAPT $\A$ such
that $\L(\A) = \L(\A_1) \cap \L(\A_2)$ and whose size is linear in
the size of $\A_1$ and $\A_2$.
\end{lemma}

We now recall a result on \GAPT and hybrid graded $\mu$-calculus
formulas.

\begin{lemma}[\cite{BLMV06}]\label{lem:FromFormulasToAutomata}
Given an hybrid graded $\mu$-calculus sentence $\varphi$ with $\ell$
\emph{atleast} subsentences and counting up to $b$, it is possible to construct
a \GAPT with $\mathcal{O}(|\varphi|^2)$ states, index $|\varphi|$, and counting
bound $b$ that accepts exactly each tree that encodes a quasi-forest model of
$\varphi$.
\end{lemma}

\subsection{Nondeterministic Pushdown Parity Tree Automata
(\PDNPT)}\label{sub:Pdnpt} A \PDNPT (without $\varepsilon$-transitions), on
$\Sigma$-labeled full $h$-ary trees, is a tuple $\P = \tpl{\Sigma, \Gamma,
\bottom, Q, q_0, \gamma_0$, $\rho,\F}$, where $\Sigma$ is a finite input
alphabet, $\Gamma$, $\bottom$, $\Gb$, and $Q$ are as in \OPD, $(q_0,\gamma_0)$
is the initial configuration, $\rho:Q\times \Sigma \times \Gb \rightarrow 2^{(Q
\times \Gb^*)^h}$ is a transition function, and $\F$ is a parity acceptance
condition over $Q$. Intuitively, when $\P$ is in state $q$, reading an input
node $x$ labeled by $\sigma \in \Sigma$, and the stack contains a word $A \cdot
\gamma \in \Gamma^* \cdot \bottom$, then $\P$ chooses a tuple
$\tpl{(q_1,\gamma_1),\ldots,(q_h,\gamma_h)} \in \rho(q,\sigma,A)$ and splits in
$h$ copies such that for each $1\leq i\leq h$, a copy in configuration $(q_i,
\gamma_i \cdot \gamma)$ is sent to the node $x \cdot i$ in the input tree. A
run of $\P$ on a $\Sigma$-labeled full $h$-ary tree $\tpl{T,V}$ is a $(Q \times
\Gamma^* \cdot \bottom)$-labeled tree $\tpl{T,r}$ such that
\begin{enumerate}[$\bullet$]

  \item
  $r(\varepsilon) = (q_0,\gamma_0)$, and

  \item
  for each $x\in T$ with $r(x)=(q,A\cdot\gamma)$, there is
  $\tpl{(q_1,\gamma_1),\ldots,(q_h,\gamma_h)} \in \rho(q,V(x),A)$
  such that, for all $1\leq i\leq h$, we have $r(x \cdot i) =
  (q_i,\gamma_i \cdot \gamma)$.

\end{enumerate}
The notion of accepting path is defined with respect to the
control states that appear infinitely often in the path (thus
without taking into account any stack content). Then, the notions
given for \TGAPT regarding accepting runs, accepted trees, and
accepted languages, along with the parity acceptance condition,
easily extend to \PDNPT. In the following, we denote with \PDNBT a
\PDNPT with a B\"uchi condition. We now recall two useful results
on the introduced automata.
\begin{prop}[\cite{KPV02}]\label{prop:EmptinessForPD-NBT}
The emptiness problem for a \PDNPT on $\Sigma$-labeled full
$h$-ary trees, having index $m$, $n$ states, and transition
function $\rho$, can be solved in time exponential in $n \cdot m
\cdot h \cdot |\rho|$.
\end{prop}
\begin{prop}[\cite{BMP05}]\label{pro:ClosureUnderIntersection}
Given a $\PDNBT$ $\P = \langle \Sigma$, $\Gamma$, $Q$, $q_0$,
$\gamma_0$, $\rho$, $Q \rangle$ on $\Sigma$-labeled full $h$-ary
trees, and an $\NPT$ $\A = \tpl{\Sigma, Q', q'_0, \delta, \F'}$,
there is a $\PDNPT$ $\P'$ on $\Sigma$-labeled full $h$-ary trees,
such that $\L(\P') = \L(\P) \cap \L(\A)$. Moreover, $\P'$ has
$|Q|\cdot |Q'|$ states, the same index as $\A$, and the size of
the transition relation is bounded by $|\rho| \cdot |\delta| \cdot
h$.
\end{prop}

\section{Deciding Hybrid Graded $\mu$-calculus Module Checking}\label{sec:DecidingHybridGradedModuleChecking}
In this section, we solve the module checking problem for the
hybrid graded $\mu$--calculus. In particular, we show that this
problem is decidable and \EXPTIME--complete. For the upper bound,
we give an algorithm based on an automata--theoretic approach, by
extending an idea of \cite{KVW01}. For the lower bound, we give a
reduction from the module checking problem for \CTL, known to be
\EXPTIME--hard. We start with the upper bound.

Let $\M$ be a module and $\varphi$ an hybrid graded
$\mu$--calculus formula. We decide the module checking problem for
$\M$ against $\varphi$ by building a \GAPT $\A_{\M \times
\not\models\varphi}$ as the intersection of two automata.
Essentially, the first automaton, denoted by $\A_{\M}$, is a
B\"{u}chi automaton that accepts trees encoding of labeled
quasi--forests of $exec(\M)$, and the second automaton is a \GAPT
$\A_{\not\models\varphi}$ that accepts all trees encoding of
labeled quasi--forests that do not satisfy $\varphi$ (i.e, $\neg
\varphi$ is satisfied at all initial nodes). Thus, $\M \models_r
\varphi$ iff $\L(\A_{\M \times \not\models\varphi})$ is empty.

The construction of $\A_{\M}$ proposed here extends that given in
\cite{KVW01} for solving the module checking problem for
finite--state open systems with respect to \CTL and \CTLSTAR. The
extension concerns the handling of forest models instead of trees
and formulas of the hybrid graded $\mu$--calculus. Before
starting, there are a few technical difficulties to be overcome.
First, we notice that $exec(\M)$ contains quasi--forests, with
labels on both edges and nodes, while B\"{u}chi automata can only
accept trees with labels on nodes. This problem is overcome by
using the following three step transformation
\begin{enumerate}[(1)]
\item
move the label of each edge to the target node of the edge (formally using a
new propositional symbol $p_{\alpha}$, for each atomic program $\alpha$),

\item
substitute edges to roots with new propositional symbols $\uparrow_o^{\alpha}$
(which represents an $\alpha$--labeled edge from the current node to the unique
root node labeled by the nominal $o$), and

\item
add a new root, labeled with a new symbol $\rr$, and connect it with the old roots of the quasi--forest.
\end{enumerate}

Let $AP'= AP \cup \{p_{\alpha} \mid \alpha \in Prog\} \cup
\{\uparrow_o^{\alpha} \mid \alpha \in Prog \text{ and } o \in Nom\}$, we denote
with $\tpl{T,\node'}$ the $(2^{AP' \cup Nom} \cup \{\rr\})$--labeled tree
encoding of a quasi--forest $\tpl{F,\node,\edge} \in exec(\M)$, obtained using
the above transformation.

Another technical difficulty to handle is relate to the fact that
quasi--forests of $exec(\M)$ (and thus their encoding) may not share the same
structure, since they are obtained by pruning some subtrees from the
computation quasi--forest $\tpl{F_{\M}, \node_{\M}, \edge_{\M}}$ of $\M$. Let
$\tpl{T_{\M}, \node_{\M}'}$ the \emph{computation tree} of $\M$ obtained from
$\tpl{F_{\M}, \node_{\M}, \edge_{\M}}$ using the above encoding. By extending
an idea of \cite{KVW01}, we solve the technical problem by considering each
tree $\tpl{T,\node'}$, encoding of a quasi--forest of $exec(\M)$, as a
$(2^{AP'\cup Nom} \cup \{\rr,\bot\})$--labeled tree $\tpl{T_{\M}, \node''}$
(where $\bot$ is a fresh proposition name not belonging to $AP \cup Nom \cup
\{\rr\}$) such that for each node $x \in T_{\M}$, if $x \in T$ then $\node''(x)
= \node'(x)$, otherwise $\node''(x) = \{\bot\}$. Thus, we label each node
pruned in the $\tpl{T_{\M},\node_{\M}'}$ with $\{\bot\}$ and recursively, we
label with $\{\bot\}$ its subtrees. In this way, all trees encoding
quasi--forests of $exec(\M)$ have the same structure of
$\tpl{T_{\M},\node_{\M}'}$, and they differ only in their labeling.

Accordingly, we can think of an environment as a strategy for placing
$\{\bot\}$ in $\tpl{T_{\M},\node_{\M}'}$, with the aim of preventing the system
to satisfy a desired property while not considering the nodes labeled with
$\bot$. Moreover, the environment can also disable jumps to roots. This is
performed by removing from nodes corresponding with environment states some of
$\uparrow_o^{\alpha}$ labels. Notice that since we consider environments that
do not block the system, each node associated with an environment state has at
least one successor not labeled by $\{\bot\}$, unless it has
$\uparrow_o^{\alpha}$ in its label.

Let us denote by $\widehat{exec}(\M)$ the set of all $(2^{AP' \cup Nom} \cup
\{\rr,\bot\})$--labeled $\tpl{T_{\M},\node''}$ trees obtained from
$\tpl{F,\node,\edge} \in exec(\M)$ in the above described manner. The required
\NBT $\A_{\M}$ must accept all and only the $(2^{AP' \cup Nom} \cup
\{\rr,\bot\})$--labeled trees in $\widehat{exec}(\M)$. The automaton
$\A_{\M}=\tpl{\Sigma, D, Q, \delta, q_0, \F}$ is defined for a module $\M =
\tpl{W_s, W_e,W_0, R, L}$  as follows:
\begin{enumerate}[$\bullet$]
    \item
    $\Sigma = 2^{AP'\cup Nom} \cup \{\rr,\bot\}$

    \item
    $D = \bigcup_{w \in W} |succ(w) \setminus W_0|$ (that is, $D$ contains, for
    each state in $W$, the number of its successors, but its jumps to roots).

    \item
    $Q = (W \times \{\bot,\top,\vdash\}) \cup \{q_0\}$, with $q_0 \not\in W$. Thus
    every state $w$ of $\M$ induces three states $(w,\bot)$, $(w,\top)$, and
    $(w,\vdash)$ in $\A_{\M}$. Intuitively, when $\A_{\M}$ is in state $(w,\bot)$,
    it can read only $\bot$, in state $(w,\top)$, it can read only letters in
    $2^{AP^*\cup\ Nom}$, and in state $(w,\vdash)$, it can read both letters in
    $2^{AP^*\cup Nom}$ and $\bot$. In this last case, it is left to the environment
    to decide whether the transition to a state of the form $(w,\vdash)$ is
    enabled. The three types of states are used to ensure that the environment
    enables all transitions from enabled system states, enables at least one
    transition from each enabled environment state, and disables transitions from
    disabled states.

    \item
    The transition function $\delta: Q \times \Sigma \times D
    \rightarrow 2^{Q^*}$ is defined as follows. Let $x \in T$ be a
    node of the input tree.
    \begin{enumerate}[$-$]
    \item
    if $\rr \in \node(x)$ then (let $W_0 = \{w_1, \ldots, w_m\}$)
        $$\delta(q_0,\rr,m) = \{ \tpl{(w_1,\top), \ldots, (w_m,\top)} \},$$
    that is $\delta(q_0,\rr,m)$ contains exactly one $m$--tuple of all
    the roots of the forest. In this case, all transitions cannot be
    disabled;

    \item
    if $\rr \not \in \node(x)$, let $\node(x) = w$ and $succ(w)
    \setminus W_0=\tpl{w_1, \ldots, w_n}$ be the set of
    \emph{non--roots successors} of $w$, then we have

\begin{enumerate}[$*$]
\item
for $w \in W_e \cup W_s$ and $g \in \{\vdash,\bot\}$ we have
    $$\delta((w,g),\bot,n) = \{ \tpl{(w_1,\bot), \ldots, (w_n,\bot)} \},$$
that is $\delta((w,g),\bot,n)$ contains exactly one $n$--tuple of
all non--roots successors of $w$. In this case, all transitions to
successors of $w$ are recursively disabled;

\item
for $w\in W_s$ and $g\in \{\top,\vdash\}$ we have
    $$\delta((w,g),L(w),n) = \{ \tpl{(w_1,\top), \ldots, (w_n,\top)} \},$$
that is, $\delta((w,g),L(w),n)$ contains exactly one $n$--tuple of
all non--roots successors of $w$. In this case all transitions to
successors of $w$ are enabled;

\item
for $w\in W_e$ and $g\in \{\top,\vdash\}$ with $L(w) \cap
\{\uparrow_o^{\alpha} \mid \alpha \in Prog \text{ and } o \in Nom
\} = \emptyset$ (i.e., $w$ has no jumps to roots or all of them
have been disabled), we have
\begin{displaymath}
    \delta((w,g),L(w),n) = \{\begin{array}[t]{l}
        \tpl{(w_1,\top),\ (w_2,\vdash),\ldots,(w_n,\vdash)},\\
        \tpl{(w_1,\vdash),\ (w_2,\top),\ldots,(w_n,\vdash)},\\
        \hspace{2cm} \vdots \\
        \tpl{(w_1,\vdash),\ (w_2,\vdash),\ldots,(w_n,\top)}\},
    \end{array}
\end{displaymath}
that is, $\delta((w,g),L(w),n)$ contains $n$ different $n$--tuples
of all non--roots successors of $w$. When $\A_{\M}$ proceeds
according to the $i$--th tuple, the environment can disable all
transitions to successors of $w$, except that to $w_i$;

\item
for $w\in W_e$ and $g\in \{\top,\vdash\}$ with $L(w) \cap
\{\uparrow_o^{\alpha} \mid \alpha \in Prog \text{ and } o \in Nom
\} \neq \emptyset$ (i.e., $w$ has at least one jump to roots
enabled), we have
    $$\delta((w,g),L(w),n) = \{ \tpl{(w_1,\vdash), \ldots, (w_n,\vdash)} \},$$
that is $\delta((w,g),L(w),n)$ contains one $n$--tuple of
non--roots successors of $w$, that can be successively disabled.
\end{enumerate}
\end{enumerate}
\end{enumerate}

Notice that $\delta$ is not defined when $n$ is different from the
number of non--roots successors of $w$, and when the input does
not meet the restriction imposed by the $\top$, $\vdash$, and
$\bot$ annotations or by the labeling of $w$.

The automaton $\A_{\M}$ has $3 \cdot |W| + 1$ states, $2^{|AP|
\cdot |R|}+2$ symbols, and the size of the transition relation
$|\delta|$ is bounded by $|R|(|W| \cdot 2^{|R|})$.

We recall that a node labeled by either $\{\bot\}$ or $\{\rr\}$
stands for a node that actually does not exist. Thus, we have to
take this into account when we interpret formulas of the hybrid
graded $\mu$--calculus over trees $\tpl{T_{\M},V'} \in
\widehat{exec}(\M)$. In order to achieve this, as in \cite{KVW01}
we define a function $f$ that transforms the input formula
$\varphi$ in a formula of the hybrid graded $\mu$--calculus
$\varphi'= \tpl{0,\alpha}f(\varphi)$ (where $\alpha \in Prog$ is
an arbitrary atomic program), that restricts path quantification
to only paths that never visit a state labeled with $\{\bot\}$.
The function $f$ we consider extends that given in \cite{KVW01}
and is inductively defined as follows:
\begin{enumerate}[$\bullet$]
\item
$f(\true) = \true$ and $f(\false) = \false$;

\item
$f(p) = p$ and $f(\neg p) = \neg p$ for all $p \in AP \cup Nom$;

\item
$f(x) = x$ for all $x \in Var$;

\item
$f(\varphi_1 \vee \varphi_2) = f(\varphi_1) \vee f(\varphi_2)$ and
$f(\varphi_1 \wedge \varphi_2) = f(\varphi_1) \wedge f(\varphi_2)$
for all hybrid graded $\mu$--calculus formulas $\varphi_1$ and
$\varphi_2$;

\item
$f(\mu x.\varphi(x))= \mu x. f(\varphi(x))$ and $f(\nu x.
\varphi(x))= \nu x. f(\varphi(x))$ for all $x \in Var$ and hybrid
graded $\mu$--calculus formulas $\varphi$;

\item
$f(\tpl{n,\aa}\varphi) = \tpl{n,\aa}(\neg \bot \wedge f(\varphi))$
for $n \in \Naturals$ and for all atomic programs $\aa$ and hybrid
graded $\mu$--calculus formulas $\varphi$;

\item
$f([n,\aa]\varphi) = [n,\aa](\neg \bot \wedge f(\varphi))$ for $n
\in \Naturals$ and for all atomic programs $\aa$ and hybrid graded
$\mu$--calculus formulas $\varphi$.
\end{enumerate}

By definition of $f$, it follows that for each formula $\varphi$
and $\tpl{T,\node} \in \widehat{exec}(\M)$, $\tpl{T,\node}$
satisfies $\varphi'=\tpl{0,\alpha}f(\varphi)$ \emph{iff} the
$2^{AP'\cup Nom}$--labeled forest, obtained from $\tpl{T,V}$
removing the node labeled with $\{\rr\}$ and all nodes labeled by
$\{\bot\}$, satisfies $\varphi$. Therefore, we solve the module
checking problem of $\M$ against an hybrid graded $\mu$--calculus
formula $\varphi$ by checking (for its negation) that in
$\widehat{exec}(\M) = \L(\A_{\M})$ does not exist any tree
$\tpl{T,\node}$ satisfying $\neg \varphi'=[0,\alpha]f(\neg
\varphi)$ (note that $|f(\neg \varphi)| =
\mathcal{O}(|\neg\varphi|)$). We reduce the latter to check the
emptiness of a \GAPT $\A_{\M \times \not\models f(\varphi)}$ that
is defined as the intersection of the \NBT $\A_{\M}$ with a \GAPT
$\A_{\not \models f(\varphi)}$ accepting exactly the $2^{AP'\cup
Nom} \cup \{\rr,\bot\}$ trees encodings of quasi--forests not
satisfying $f(\varphi)$. By Lemma
\ref{lem:FromFormulasToAutomata}, if $\varphi$ is an hybrid graded
$\mu$--calculus formula, then $\A_{\not \models f(\varphi)}$ has
$\mathcal{O}(|\varphi|^2)$ states, index $|\varphi|$, and counting
bound $b$. Therefore, by Lemma \ref{lem:TgaptIntersection},
$\A_{\M\times\not\models f(\varphi)}$ has
$\mathcal{O}(|W|+|\varphi|^2)$ states, index $|\varphi|$, and
counting bound $b$. By recalling that the emptiness problem for a
\GAPT can be decided in exponential-time (Theorem
\ref{the:TgaptEmptiness}), we obtain that the module checking
problem for hybrid graded $\mu$--calculus formulas is solvable in
exponential-time. To show a tight lower bound we recall that \CTL
module checking is \EXPTIME--hard \cite{KVW01} and every \CTL
formula can be linearly transformed in a modal $\mu$--calculus
formula \cite{Jur98}. This leads to the module checking problem
w.r.t. modal $\mu$--calculus formulas to be \EXPTIME--hard and
thus to the following result.

\begin{theorem}\label{the:HybridGradedFiniteStatesModuleChecking}
The module checking problem with respect to hybrid graded
$\mu$--calculus formulas is \EXPTIME--complete.
\end{theorem}

\section{Deciding Hybrid Graded $\mu$-calculus PD-module Checking}\label{sec:DecidingHybridGradedPushdownModuleChecking}
In this section, we show that hybrid graded pushdown module checking is
decidable and solvable in \TWOEXPTIME. Since \CTL pushdown module checking is
\TWOEXPTIME--hard, we get that the addressed problem is \TWOEXPTIME--complete.
For the upper bound, the algorithm works as follows. Given an \OPD $\S$ and the
module $\M_{\S}$ induced by $\S$, by combining and extending the constructions
given in \cite{BMP05} and Section \ref{sec:DecidingHybridGradedModuleChecking},
we first build in polynomial--time a \PDNBT $\A_{\S}$ accepting each tree that
encodes a quasi--forest belonging to $exec(\M_{\S})$. Then, given an hybrid
graded $\mu$--calculus formula $\varphi$, according to \cite{BLMV06}, we build
in polynomial--time a \GAPT $\A_{\not\models\varphi}$ (Lemma
\ref{lem:FromFormulasToAutomata}) accepting all models that do not satisfy
$\varphi$, with the intent of checking that none of these models are in
$exec(\M_{\S})$. Then, accordingly to the basic idea of \cite{KVW01}, we check
that $\M_{\S} \models_r \varphi$ by checking whether $\L(\A_{\S}) \cap
\L(\A_{\not\models\varphi})$ is empty. Finally, we get the result by using an
exponential--time reduction of the latter to the emptiness problem for \PDNPT,
which from Proposition~\ref{prop:EmptinessForPD-NBT} can be solved in \EXPTIME.
As a key step of the above reduction, we use the exponential--time translation
from \GAPT into \NPT\ showed in Lemma~\ref{lem:FromTgaptToNpt}.

Let us start dealing with $\A_{\S}$. Before building the
automaton, there are some technical difficulties to overcome.
First, notice that $\A_{\S}$ is a \PDNBT and it can only deal with
trees having labels on nodes. Also, quasi--forests of
$exec(\M_{\S})$ may not share the same structure, since they are
obtained by pruning subtrees from the computation quasi--forest
$\tpl{F_{\M_{\S}},\node_{\M_{\S}}, \edge_{\M_{\S}}}$ of $\M_{\S}$.
As in Section \ref{sec:DecidingHybridGradedModuleChecking}, we
solve this problem by considering  $2^{\mathit{AP}' \cup Nom}
\cup\{\rr, \bot\}$--labeled trees \emph{encoding} of
quasi--forests $\tpl{F,\node,\edge} \in exec(\M_{\S})$, where
$AP'= AP \cup \{p_{\alpha} \mid \alpha \in Prog\} \cup
\{\uparrow_o^{\alpha} \mid \alpha \in Prog \text{ and } o \in
Nom\}$.

Another technical difficulty to handle with is related to the fact
that quasi--forests of $exec(\M_{\S})$ (and thus their encodings)
may not be full $h$--ary, since the nodes of the \OPD from which
$\M_{\S}$ is induced may have different degrees. Technically, we
need this property since the emptiness problem for \PDNPT to which
we reduce our problem has been solved in the literature only for
\PDNPT working on full trees. Similarly as we did for pruned
nodes, we transform each tree encoding of a quasi--forest of
$exec(\M_{\S})$ into a full $h$--ary tree by adding missing nodes
labeled with $\{\bot\}$.
Therefore the proposition $\bot$ is used to denote both
``disabled" states and ``completion" states. In this way, all
trees encodings of quasi--forests of $exec(\M_{\S})$ are all full
$h$--ary trees, and they differ only in their labeling. Let us
denote with $\widehat{exec}(\M_{\S})$ the set of all $2^{AP' \cup
Nom} \cup \{\rr, \bot\}$--labeled full $h$--ary trees obtained
from $\tpl{F_{\M_{\S}},\node_{\M_{\S}}, \edge_{\M_{\S}}}$ using
all the transformations described above.

In \cite{BMP05} it has been shown how to build a \PDNBT accepting
full $h$--ary trees embedded in an \OPD corresponding to all
behaviors of the environment. In particular, the \PDNBT
constructed there already takes into account the above
transformation regarding $\{\bot\}$--labeled nodes. By extending
the construction proposed there in the same way the construction
showed in Section \ref{sec:DecidingHybridGradedModuleChecking}
extends the classical construction of $\A_{\M}$ proposed in
\cite{KVW01}, it is not hard to show that the following result
holds.

\begin{lemma}\label{lem:FromOpdToAutomata}
Given an \OPD $\S = \tpl{Q, \Gamma, \bottom, C_0, \Delta, \rho_1,
\rho_2, Env}$ with branching degree $h$, we can build a \PDNBT
$\A_{\S} = \tpl{\Sigma, \Gamma, \bottom, Q', q_0', \gamma_0,
\delta, Q}$, which accepts exactly $\widehat{exec}(\M_{\S})$, such
that $\Sigma = 2^{AP' \cup Nom} \cup \{\rr,\bot\}$, $|Q'| =
\mathcal{O}(|Q|^2 \cdot |\Gamma|)$, and $|\delta|$ is polynomially
bounded by $h \cdot |\Delta|$.
\end{lemma}

Let us now go back to the hybrid graded $\mu$--calculus formula
$\varphi$. Using the function $f$ introduced in
Section~\ref{sec:DecidingHybridGradedModuleChecking} and
Lemma~\ref{lem:FromFormulasToAutomata}, we get that given an
hybrid graded $\mu$--calculus formula $\varphi$, we can build in
polynomial--time a \GAPT $\A_{\not \models f(\varphi)}$ accepting
all models of $\neg\varphi' = [0,\alpha]f(\neg\varphi)$ (as done
in Section \ref{sec:DecidingHybridGradedModuleChecking}).

By using the classical \EXPTIME transformation from \GAPT to \GNPT
\cite{KSV02} and a simple \EXPTIME transformation from \GNPT to
\NPT, we directly get a \THREEEXPTIME algorithm for the hybrid
graded $\mu$--calculus pushdown module checking. To obtain an
exponential--time improvement, here we show a not trivial \EXPTIME
transformation from \TGAPT to \NPT.
The translation we propose uses the notions of \emph{strategies},
\emph{promises} and \emph{annotations}, which we now recall.

Let $\A = \tpl{\Sigma, b, Q, \delta, q_0, \F}$ be a \TGAPT with
$\F = \tpl{F_1, \ldots, F_k}$ and $\tpl{T,V}$ be a
$\Sigma$-labeled tree. Recall that $D_b = \tpl{[b]} \cup [[b]]
\cup \{-1, \varepsilon\}$ and $\delta: (Q \times \Sigma)
\rightarrow B^{+}(D_b \times Q)$. For each control state $q \in
Q$, let $index(q)$ be the minimal $i$ such that $q \in F_i$. A
\emph{strategy tree} for $\A$ on $\tpl{T,V}$ is a $2^{Q \times D_b
\times Q}$-labeled tree $\tpl{T,\strat}$ such that, defined
$\head(w) = \{q : (q, d, q') \in w \}$ as the set of
\emph{sources} of $w$, it holds that $(i)$ $q_0 \in
\head(\strat(\mathit{root}(T)))$ and $(ii)$ for each node $x \in
T$ and state $q$, the set $\{(q, q') : (q,d,q') \in \strat(x)\}$
satisfies $\delta(q,V(x))$.

A \emph{promise tree} for $\A$ on $\tpl{T,V}$ is a $2^{Q \times
Q}$-labeled tree $\tpl{T,\prom}$. We say that $\prom$
\emph{fulfills} $\strat$ for $V$ if the states promised to be
visited by $\prom$ satisfy the obligations induced by $\strat$ as
it runs on $V$. Formally, $\prom$ fulfills $\strat$ for $V$ if for
every node $x \in T$, the following hold: ``for every $(q,
\tpl{n}, q') \in \strat(x)$ (resp. $(q, [n], q') \in \strat(x)$),
at least $n+1$ (resp $deg(x)-n$) successors $x \cdot j$ of $x$
have $(q,q') \in \prom(x \cdot j)$''.

An \emph{annotation tree} for $\A$ on $\tpl{T,\strat}$ and $\tpl{T, \prom}$ is
a $2^{Q \times \{1, \ldots, k\} \times Q}$-labeled tree $\tpl{T,\ann}$ such
that for each $x \in T$ and $(q, d_1, q_1) \in \strat(x)$ the following hold:

\begin{enumerate}[$\bullet$]

\item
if $d_1 = \varepsilon$, then $(q, index(q_1), q_1) \in \ann(x)$;

\item
if $d_1 \in \{1, \ldots, k\}$, then for all $d_2 \in \{1, \ldots,
k\}$ and $q_2 \in Q$ such that $(q_1, d_2, q_2) \in \ann(x)$, we
have $(q, \min(d_1,d_2), q_2) \in \ann(x)$;

\item
if $d_1 = -1$ and $x = y \cdot i$, then for all $d_2, d_3 \in \{1,
\ldots, k\}$ and all $q_2, q_3 \in Q$ satisfing $(q_1, d_2, q_2) \in
\ann(y)$ as well as $(q_2, d_3, q_3) \in \strat(y)$ and $(q_2, q_3) \in
\prom(x)$, we have that $(t, \min(index(q_1), d_2, index(q_3)),
q_3) \in \ann(x)$;

\item
if $d_1 \in [[b]] \cup \tpl{[b]}$, $y = x \cdot i$, and $(q, q_1)
\in \prom(y)$, then for all $d_2, d_3 \in \{1, \ldots, k\}$ and
$q_2, q_3 \in Q$ such that $(q_1, d_2, q_2) \in \ann(y)$ and
$(q_2, -1, q_3) \in \strat(y)$, it holds that $(t,
\min(index(q_1), d_2, index(q_3)), q_3) \in \ann(x)$.
\end{enumerate}

A downward path induced by $\strat$, $\prom$, and $\ann$ on
$\tpl{T,V}$ is a sequence $\tpl{x_0,q_0,t_0}$,
$\tpl{x_1,q_1,t_1}$, $\ldots$ such that $x_0 = root(T)$, $q_0$ is
the initial state of $\A$ and, for each $i \geq 0$, it holds that
$x_i \in T$, $q_i \in Q$, and $t_i = \tpl{q_i, d, q_{i+1}} \in
\strat(x_i) \cup \ann(x_i)$ is such that either \emph{(i)} $d \in
\{1, \ldots, k\}$ and $x_{i+1} = x_i$, or \emph{(ii)} $d \in
\tpl{[b]} \cup [[b]]$ and there exists $c \in \{1, \ldots,
\degree(x_i)\}$ such that $x_{i+1} = x_i \cdot c$ and
$(q_i,q_{i+1}) \in \prom(x_{i+1})$. In the first case we set
$index(t_i) = d$ and in the second case we set $index(t_i) =
\min\{j \in \{1, \ldots, k\} \mid q_{i+1} \in F_j\}$. Moreover,
for a downward path $\pi$, we set $index(\pi)$ as the minimum
index that appears infinitely often in $\pi$. Finally, we say that
$\pi$ is \emph{accepting} if $index(\pi)$ is even.

The following lemma relates languages accepted by \TGAPT with
strategies, promises, and annotations.
\begin{lemma}[\cite{BLMV06}]\label{lem:TgaptAndStrategies}
Let $\A$ be a \TGAPT. A $\Sigma$-labeled tree $\tpl{T,V}$ is
accepted by $\A$ iff there exist a strategy tree $\tpl{T,\strat}$,
a promise tree $\tpl{T,\prom}$ for $\A$ on $\tpl{T,V}$ such that
$\prom$ fulfills $\strat$ for $V$, and an annotation tree
$\tpl{T,\ann}$ for $\A$ on $\tpl{T,V}$, $\tpl{T,\strat}$ and
$\tpl{T,\prom}$ such that every downward path induced by $\strat$,
$\prom$, and $\ann$ on $\tpl{T,V}$ is accepting.
\end{lemma}

Given an alphabet $\Sigma$ for the input tree of a \TGAPT with transition
function $\delta$, let $D_b^{\delta}$ be the subset containing only the
elements of $D_b$ appearing in $\delta$. Then we denote by $\Sigma'$ the
extended alphabet for the combined trees, i.e., $\Sigma' = \Sigma \times 2^{Q
\times D_b^{\delta} \times Q} \times 2^{Q \times Q} \times 2^{Q \times
\{1,\ldots k\} \times Q}$.

\begin{lemma}\label{lem:FromTgaptToNpt}
Let $\A$ be a \TGAPT running on $\Sigma$--labeled trees with $n$
states, index $k$ and counting bound $b$ that accepts $h$-ary
trees. It is possible to construct in exponential-time an \NPT
$\A'$ running on $\Sigma'$--labeled $h$-ary trees that accepts a
tree iff $\A$ accepts its projection on $\Sigma$.
\end{lemma}

\begin{proof}
Let $\A = \tpl{\Sigma, b, Q, q_0, \delta, \F}$ with $\F =
\tpl{F_1, \ldots, F_k}$. By Lemma \ref{lem:TgaptAndStrategies}, we
construct $\A'$ as the intersection of two \NBT $\A'$, $\A''$, and
an \NPT $\A'''$. In particular, all these automata have size
exponential in the size of $\A$. Moreover, since each \NBT uses as
accepting all its states, it is easy to intersect in
polynomial-time all of them by using a classical automata product.
These automata are defined as follows. Given a $\Sigma'$-labeled
tree $T' = \tpl{T,(V,\strat,\prom,\ann)}$,
\begin{enumerate}[(1)]
\item
$\A'$ accepts $T'$ iff $\strat$ is a strategy for $\A$ on
$\tpl{T,V}$ and $\prom$ fulfills $\strat$ for $V$,

\item
$\A''$ accepts $T'$ iff $\ann$ is an annotation for $\A$ on
$\tpl{T,V}$, $\tpl{T,\strat}$ and $\tpl{T,\prom}$, and

\item
$\A'''$ accepts $T'$ iff every downward path induced by $\strat$,
$\prom$, and $\ann$ on $\tpl{T,V}$ is accepting.
\end{enumerate}

The automaton $\A' = \tpl{\Sigma', D', Q', q'_0, \delta', \F'}$
works as follows: on reading a node $x$ labeled $(\sigma, \eta,
\rho, \omega)$, then it locally checks whether $\eta$ satisfies
the definition of strategy for $\A$ on $\tpl{T,V}$. In particular,
when $\A'$ is in its initial state, we check that $\eta$ contains
a transition starting from the initial state of $\A$. Moreover,
the automaton $\A'$ sends to each child $x \cdot i$ the pairs of
states that have to be contained in $\prom(x \cdot i)$, in order
to verify that $\prom$ fulfills $\strat$. To obtain this, we set
$Q' = 2^{Q \times Q} \cup \{q'_0\}$, $D' = \{1, \ldots, h\}$ and
$\F' = \{\emptyset, Q'\}$. To define $\delta'$, we first give the
following definition. For each node $x \in T$ labeled $(\sigma,
\eta, \rho, \omega)$, we set {\small
    $$\begin{array}{lcl}
        S(\eta) &=& \{\tpl{S_1, \ldots, S_{\degree(x)}} \in (2^{Q \times Q})^{\degree(x)} \mbox{ such that} \\
        & & [\mbox{for each } (q, \tpl{m}, p) \in \eta\ \mbox{ there is } P \subseteq \{1, \ldots \degree(x)\}
            \mbox{ with } |P| = m+1 \\
        & & \mbox{such that for all } i \in P,\ (q,p) \in S_i] \mbox{ and} \\
        & & [\mbox{for each } (q, [m], p) \in \eta\ \mbox{ there is}  P \subseteq \{1, \ldots \degree(x)\}
            \mbox{ with } |P| = \degree(x)-m \\
        & & \mbox{such that for all } i \in P,\ (q,p) \in S_i]\}\
    \end{array}$$}
to be the set of all tuples with size $\degree(x)$, each
\emph{fulfilling} all graded modalities in $\strat(x)$. Notice
that $|S(\eta)| \leq 2^{hn^2}$. Then we have {\small
    $$\delta'(q, (\sigma, \eta, \rho, \omega), \degree(x)) = \begin{cases}
        S(\eta)     & \mbox{if } \forall\ p \in head(\eta),\ \{(d,p') \mid (p,d,p') \in \eta\}
                              \mbox{ satisfies } \delta(p, \sigma) \\
                    & \mbox{and } [(q = q_0^1 \mbox{ and } q_0 \in head(\eta))
                              \mbox{ or } (q \neq q_0^1 \mbox{ and } q \subseteq \rho)] \\
        {\bf false} & \mbox{otherwise.}
    \end{cases}$$}
\noindent Hence, in $\A'$ we have $|Q'| = 2^{n^2}$, $|\delta'|
\leq 2^{n^2(k+1)}$, and index $2$.

$\A'' = \tpl{\Sigma', D'', Q'', q''_0, \delta'', \F''}$ works in a
similar way to $\A'$. That is, for each node $x$, it first locally
checks whether the constraints of the annotations are verified;
then it sends to the children of $x$ the strategy and annotation
associated with $x$, in order to successively verify whether the
promises associated with the children nodes are consistent with
the annotation of $x$. Therefore, in $A''$ we have $Q'' = 2^{Q
\times D_b^{\delta} \times Q} \times 2^{Q \times \{1, \ldots, k\}
\times Q}$, $q''_0 = (\emptyset, \emptyset)$, $\F'' = \{\emptyset,
Q''\}$, $D'' = \{1, \ldots, h\}$, and for a state $(\eta_{prev},
\omega_{prev})$ and a letter $(\sigma, \eta, \rho, \omega)$ we
have {\small
    $$\delta''((\eta_{prev}, \omega_{prev}), (\sigma, \eta, \rho, \omega), \degree(x)) = \begin{cases}
        \tpl{(\eta,\omega), \ldots, (\eta,\omega)} & \mbox{if the local conditions for the} \\
                                                   & \mbox{annotations are verified} \\
        {\bf false}                                & \mbox{otherwise.}
    \end{cases}$$}
Hence, in $\A''$ we have $|Q''| \leq 2^{n^2(|\delta|+k)}$,
$|\delta''| \leq h \cdot 2^{n^2(|\delta|+k)}$, and index $2$.

Finally, to define $\A'''$ we start by constructing a \TAPT $\B$
whose size is polynomial in the size of $\A$ and accepts
$\tpl{T,(V,\strat,\prom,\ann)}$ iff there is a non accepting
downward path (w.r.t. $\A$) induced by $\strat$, $\prom$, and
$\ann$ on $\tpl{T,V}$. The automaton $\B = \tpl{\Sigma', Q^B,
q_0^B, \delta^B, \F^B}$ (which in particular does not need
direction $-1$) essentially chooses, in each state, the downward
path to walk on, and uses an integer to store the index of the
state. We use a special state $\sharp$ not belonging to $Q$ to
indicate that $\B$ proceeds in accordance with an annotation
instead of a strategy. Therefore, $Q^B = ((Q \cup \{\sharp\})
\times \{1, \ldots, k\} \times Q) \cup \{q_0^B\}$.

To define the transition function on a node $x$, let us introduce
a function $f$ that for each $q \in Q$, strategy $\eta \in 2^{Q
\times D_b^{\delta} \times Q}$, and annotation $\omega \in 2^{Q
\times \{1, \ldots, k\} \times Q}$ gives a formula satisfied along
downward paths consistent with $\eta$ and $\omega$, starting from
a node reachable in $\A$ with the state $q$. That is, in each node
$x$, the function $f$ either proceeds according to the annotation
$\omega$ or the strategy $\eta$ (note that $f$ does not check that
the downward path is consistent with any promise). Formally, $f$
is defined as follows, where $index(p)$ is the minimum $i$ such
that $p \in F_i$: {\small
    $$f(q, \eta, \omega) = \mathop{\bigvee_{(q, d, p) \in \omega}}_{d \in \{1, \ldots, k\}}
        \tpl{\varepsilon, (\sharp, d, p)} \hspace{1em} \vee \hspace{1em}
        \mathop{\bigvee_{(q, d, p) \in \eta}}_{d \in \tpl{[b]} \cup [[b]]}
        \bigvee_{c \in \{1, \ldots, \degree(x)\}} \tpl{c, (q, index(p), p)}$$}

Then, we have $\delta^B(q_0^B, (\sigma, \eta, \rho, \omega)) =
f(q_0, \eta, \omega)$ and

{\small
\begin{eqnarray*}
    \delta^B((q, d, p), (\sigma, \eta, \rho, \omega)) &=& \begin{cases}
        {\bf false} & \mbox{ if } q \neq \sharp \mbox{ and } (q, p) \not\in \rho \\
        f(p, \eta, \omega) & \mbox{ otherwise.}
    \end{cases}.
\end{eqnarray*}}
\noindent A downward path $\pi$ is non accepting for $\A$ if the
minimum index that appears infinitely often in $\pi$ is odd.
Therefore, $\F^B = \tpl{F_1^B, \ldots, F_{k+1}^B, Q^B}$ where
$F_1^B = \emptyset$ and, for all $i \in \{2, \ldots, k+1\}$, we
have $F_i^B = \{(q,d,p) \in Q^B \mid d = i-1\}$. Thus,
$|Q^B|=kn(n+1)+1$, $|\delta^B| =k \cdot |\delta| \cdot |Q^B|$, and
the index is $k+2$. Then, since $\B$ is alternating, we can easily
complement it in polynomial-time into a \TAPT $\overline{\B}$ that
accepts a tree iff all downward paths induced by $\strat$,
$\prom$, and $\ann$ on $\tpl{T,V}$ are accepting. Finally,
following \cite{Var98} we construct in exponential-time the
desired automaton $\A'''$.
\end{proof}

By applying the transformation given by Lemma
\ref{lem:FromTgaptToNpt} to the automaton $\A_{\neg\varphi'}$
defined above, we obtain in exponential time in the size of
$\varphi$, an \NPT that accepts all the trees encoding of
quasi--forests that do not satisfy $\varphi$. From Proposition
\ref{pro:ClosureUnderIntersection}, then we can build a \PDNPT
$\A_{\S \times \not\models\varphi}$ with size polynomial in the
size of $\S$ and exponential in the size of $\varphi$ such that
$\L(\A_{\S \times \not\models\varphi}) = \L(\A_{\S}) \cap
\L(\A_{\neg\varphi'})$. Hence, from Proposition
\ref{the:TgaptEmptiness} we obtain that hybrid graded
$\mu$--calculus pushdown module checking can be solved in \EXPTIME
in the size of $\S$ and in \TWOEXPTIME in the size of $\varphi$.
Finally, from the fact that \CTL pushdown module checking is known
to be \TWOEXPTIME--hard with respect to the size of $\varphi$ and
\EXPTIME--hard with respect to the size of $\S$ \cite{BMP05}, we
obtain the following theorem.

\begin{theorem}\label{the:HybridGradedPushdownModuleChecking}
The hybrid graded $\mu$--calculus pushdown module checking problem
is \TWOEXPTIME--complete with respect to the size of the formula
and \EXPTIME--complete with respect to the size of the system.
\end{theorem}

\section{Fully Enriched $\mu$-calculus Module Checking}\label{sec:FullyEnrichedModuleChecking}
In this section, we consider a memoryless restriction of the
module checking problem and investigate it with respect to
formulas of the Fully enriched $\mu$-calculus. Given a formula
$\varphi$, a \emph{memoryless module checking} problem checks
whether all trees in $exec(\M)$, always taking the same choice in
duplicate environment nodes, satisfy $\varphi$. In this section,
we show that the (memoryless) module checking problem for Fully
enriched $\mu$-calculus is undecidable.

Fully enriched $\mu$--calculus is the extension of hybrid graded
$\mu$--calculus with \emph{inverse programs}. Essentially, inverse
programs allow us to specify properties about predecessors of a
state. Given an atomic program $a \in Prog$, we denote its inverse
program with $a^-$ and the syntax of the fully enriched
$\mu$--calculus is simply obtained from the one we introduced for
hybrid graded $\mu$--calculus, by allowing both atomic and inverse
programs in the graded modalities. Similarly, the semantics of
fully enriched $\mu$--calculus is given, identically to the one
for hybrid graded $\mu$--calculus, with respect to a Kripke
structure $\K = \tpl{W, W_0, R, L}$ in which, to deal with inverse
programs, we define, for all $a \in Prog$, $R(a^-) = \{(v,w) \in W
\times W$ such that $(w,v) \in R(a)\}$.

Let us note that, since the fully enriched $\mu$--calculus does
not enjoy the forest model property \cite{BP04}, we cannot unwind
a Kripke structure in a forest. However, it is always possible to
unwind it in an equivalent acyclic graph that we call
\emph{computation graph}. In order to take into account all the
possible behaviors of the environment, we consider all the
possible subgraphs of the computation graph obtained disabling
some transitions from environment nodes but one. We denote with
$graphs(M)$ the set of this graphs. Given a Fully enriched
$\mu$--calculus formula $\varphi$, we have that $M \models_r
\varphi$ iff $K \models \varphi$ for all $K \in graphs(M)$.

To show the undecidability of the addressed problem, we need some
further definitions. An (infinite) \emph{grid} is a tuple $G =
\tpl{\Naturals^2, h, v}$ such that $h$ and $v$ are defined as
$h(\tpl{x,y}) = \tpl{x+1,y}$ and $v(\tpl{x,y}) = \tpl{x,y+1}$.
Given a finite set of \emph{types} $T$, we will call \emph{tile}
on $T$ a function $\hat{\rho} : \Naturals^2 \rightarrow T$ that
associates a type from $T$ to each vertex of an infinite grid $G$,
and we call \emph{tiled infinite grid} the tuple
$\tpl{G,T,\hat{\rho}}$. A {\em grid model} is an infinite Kripke
structure $K = \tpl{W,\{w_0\},R,L}$, on the set of atomic programs
$Prog = \{l^-,v\}$, such that $K$ can be mapped on a grid in such
a way that $w_0$ corresponds to the vertex $\tpl{0,0}$, $R(v)$
corresponds to $v$ and $R(l^-)$ corresponds to $h$. We say that a
grid model $K$ ``corresponds'' to a tiled infinite grid
$\tpl{G,T,\hat{\rho}}$ if every state of $K$ is labeled with only
one atomic proposition (and zero or more nominals) and there
exists a bijective function $\rho : T \rightarrow AP$ such that,
if $w_{x,y}$ is the state of $K$ corresponding with the node
$\tpl{x,y}$ of $G$, then $\rho(\hat{\rho}(\tpl{x,y})) \in
L(w_{x,y})$.

\begin{theorem}\label{the:undecidability}
The module checking problem for fully enriched $\mu$--calculus is
undecidable.
\end{theorem}

\begin{proof}
To show the result, we use a reduction from the tiling problem (also known as
\emph{domino} problem), known to be undecidable \cite{Ber66}. The tiling
problem is defined as follows.

Let $T$ be a finite set of types, and $H,V \subseteq T^2$ be two relations
describing the types that cannot be vertically and horizontally adjacent in an
infinite grid. The tiling problem is to decide whether there exists a tiled
infinite grid $\tpl{G,T,\hat{\rho}}$ such that $\hat{\rho}$ preserves the
relations $H$ and $V$. We call such a tile function a \emph{legal tile} for $G$
on $T$.

In \cite{BP04}, Bonatti and Peron showed undecidability for the satisfiability
problem for fully enriched $\mu$--calculus by also using a reduction from the
tiling problem. Hence, given a set of types $T$ and relations $H$ and $V$, they
build a (alternation free) fully enriched $\mu$--calculus formula $\varphi$
such that $\varphi$ is satisfiable iff the tiling problem has a solution in a
tiled infinite grid, with a legal tile $\rho$ on $T$ (with respect to $H$ and
$V$). In particular, the formula they build can be only satisfiable on a grid
model $K$ corresponding to a tiled infinite grid with a legal tile $\rho$ on
$T$. In the reduction we propose here, we use the formula $\varphi$ used in
\cite{BP04}. It remains to define the module.

Let $\{G_1, G_2, \ldots\}$ be the set of all the infinite tiled grids on $T$
(i.e., $G_i = \tpl{G,T,\hat{\rho}_i}$), we build a module $M$ such that
$graphs(M)$ contains, for each $i\geq 1$, a grid models corresponding to $G_i$.
Therefore, we can decide the tiling problem by checking whether $M \models_r
\neg\varphi$. Indeed, if $M \models_r \neg\varphi$, then all grid models
corresponding to $G_i$ do not satisfy $\varphi$ and, therefore, there is no
solution for the tiling problem. On the other side, if $M \not\models_r
\neg\varphi$, then there exists a model for $\varphi$; since $\varphi$ can be
satisfied only on a grid model corresponding to a tiled infinite grid with a
legal tile on $T$ with respect to $H$ and $V$, we have that the tiling problem
has a solution.

Formally, let $T=\{t_1, \ldots, t_m\}$ be the set of types, the module $M =
\tpl{W_s,W_e, \allowbreak W_0,R,L}$ with respect to atomic programs $Prog =
\{l^-,v\}$, atomic propositions $AP = T$, and nominals $Nom = \{o_1, \ldots,
o_m\}$, is defined as follows:
\begin{enumerate}[$\bullet$]
\item
$W_s = \emptyset$, $W_e = \{x_1, \ldots, x_m, y_1, \ldots, y_m\}$
and $W_0 = \{x_1, \ldots, x_m\}$;

\item
for all $i \in \{1, \ldots m\}$, $L(t_i) = \{x_i, y_i\}$ and
$L(o_i) = \{x_i\}$;

\item
$R(v) = \{\tpl{x_i, x_j} | i,j \in \{1, \ldots, m\}\} \cup
\{\tpl{y_i, y_j} | i,j \in \{1, \ldots, m\}\}$;

\item
$R(l^-) = \{\tpl{x_i, y_j} | i,j \in \{1, \ldots, m\}\} \cup
\{\tpl{y_i, x_j} | i,j \in \{1, \ldots, m\}\}$
\end{enumerate}

Notice that we duplicate the set of nodes labeled with tiles since
we cannot have pairs of nodes in $M$ labeled with more than one
atomic program (in our case, with both $v$ and $l^-$). Moreover
the choice of labeling nodes $x_i$ with nominals is arbitrary.
Finally, from the fact that the module contains only environment
nodes, it immediately follows that, for each $i$, the grid model
corresponding to the infinite tiled grid $G_i$ is contained in
$graphs(M)$.
\end{proof}

\section{Notes on Fully Enriched $\mu$-calculus Model Checking}\label{sec:ModelChecking}
In this section, for the sake of completeness, we investigate the model
checking problems for Fully enriched $\mu$-calculus and its fragments, for both
pushdown and finite-states systems.

In particular, we first consider the model checking problem for formulas of the
$\mu$--calculus enriched with nominal (\emph{hybrid $\mu$--calculus}) or graded
modalities (\emph{graded $\mu$--calculus}) or both, for pushdown systems
(\PDMC, for short) and finite states systems (\FSMC, for short), i.e. Kripke
structures. In particular, we show that for graded $\mu$--calculus, \PDMC is
solvable in \TWOEXPTIME and \FSMC is solvable in \EXPTIME. Moreover we show
that for hybrid $\mu$--calculus \PDMC is \EXPTIME--complete and \FSMC is in \UP
$\cap$ \coUP, thus matching the known results for (propositional)
$\mu$--calculus model checking (see \cite{Wal96} for \PDMC and \cite{Wil01} for
\FSMC), and that, for hybrid graded $\mu$--calculus, \PDMC is solvable in
\TWOEXPTIME and \FSMC is solvable in \EXPTIME.

By considering also $\mu$-calculus enriched (among the others) with inverse
programs, we also consider \PDMC w.r.t. a reduced pushdown system that, in each
transition, can increase the size of the stack by at most one
(\emph{single--push system}). To this aim, we define a single--push system with
three stack operations: for $A \in \Gamma$, $sub(A)$ changes the top of the
stack into $A$, $push(A)$ pushes the symbol $A$ on the top of the stack, and
$pop()$ pops the top symbol of the stack. Formally, a single--push system $\S$
is a pushdown system in which the transition function is $\Delta : Prog
\rightarrow 2^{(Q \times \Gb) \times (Q \times \{sub,push,pop\} \times
\Gamma)}$. For consistency reasons, we assume that if the top of the stack is
$\bottom$ then $sub(A) = push(A)$ and $pop()$ has no effect.

We call the model checking problem for single--push systems \emph{single--push
model checking} (\SPMC, for short). In this case, we show that for full hybrid
$\mu$--calculus ($\mu$-calculus enriched with inverse programs and nominals),
\SPMC is \EXPTIME--complete and \FSMC is in \UP $\cap$ \coUP, and that for
Fully enriched and full graded $\mu$--calculus ($\mu$-calculus enriched with
inverse programs and graded modalities), \SPMC is in \TWOEXPTIME and \FSMC is
in \EXPTIME. In Figure~\ref{tab:ModelChecking} we report known and new results
on model checking problems for the Fully enriched $\mu$--calculus and its
fragments.

\begin{figure}[t]
    {\small
    \begin{tabular}{|l|c|c|c|}
        \hline
        $\mu$-calculus & Pushdown                        & Single--Push         & Finite--State                 \\
        extensions     & Model Checking                  & Model Checking       & Model Checking                \\
        \hline
        propositional  & \EXPTIME--Complete \cite{Wal96} & \EXPTIME--Complete   & \UP $\cap$ \coUP \cite{Wil01} \\
        hybrid         & \EXPTIME--Complete              & \EXPTIME--Complete   & \UP $\cap$ \coUP              \\
        graded         & \TWOEXPTIME                     & \TWOEXPTIME          & \EXPTIME                      \\
        full           & ?                               & \EXPTIME--Complete   & \UP $\cap$ \coUP              \\
        hybrid graded  & \TWOEXPTIME                     & \TWOEXPTIME          & \EXPTIME                      \\
        full hybrid    & ?                               & \EXPTIME--Complete   & \UP $\cap$ \coUP              \\
        full graded    & ?                               & \TWOEXPTIME          & \EXPTIME                      \\
        Fully enriched & ?                               & \TWOEXPTIME          & \EXPTIME                      \\
        \hline
    \end{tabular}
    \caption{Results on Model Checking Problem.}
    }
    \label{tab:ModelChecking}
\end{figure}

To prove our results, we simply rule out inverse programs and nominals from the
input formula. In particular, we first observe that, from a model checking
point of view, checking a formula with inverse programs on a graph (finite or
infinite) is equivalent to check the formula in ``forward'' on the graph
enriched with opposite edges. That is, we consider inverse programs in the
formula as special atomic programs to be checked on the opposite edges we have
added in the graph. Note that this observation does not apply to \PDMC. Indeed,
to transform previous configurations to inverse next configurations, we need to
limit the power of a \PDMC to be single push. Thus, we obtain the following
result.

\begin{lemma}\label{lem:RemovingInversePrograms}
Let $X_{\mu}$ be an enrichment of the $\mu$-calculus with inverse programs.
Then a \SPMC (resp., \FSMC) w.r.t. $X_{\mu}$ can be reduced in linear time to
\SPMC (resp., \FSMC) w.r.t. $X_{\mu}$ without inverse programs.
\end{lemma}

\begin{proof}
Here we only show the proof for \FSMC since the one for \SPMC is similar. Let
$\K = \langle W, W_0, R, L \rangle$ be a model that uses atomic programs from
$Prog$, and let $\varphi$ be a formula of $X_{\mu}$. Then, we define a new
model $\K'$ and a new formula $\varphi'$ as follows: $\K' = \langle W, W_0, R',
L \rangle$ uses atomic programs from the set $Prog' = Prog \cup \{\hat{a}$ s.t.
$a \in Prog\}$ (it doesn't use inverse programs) and has the transition
relation defined as $R'(a) = R(a)$ and $R'(\hat{a}) = R(a^-)$ for all $a \in
Prog$. On the other side, $\varphi'$ is a formula of $X_{\mu}$ without inverse
programs equal to $\varphi$ except for the fact that $a^-$ is changed into
$\hat{a}$ for all $a \in Prog$. Thus it can be easily seen that $\K \models
\varphi$ iff $\K' \models \varphi'$ and this completes the proof of this lemma.
\end{proof}

Furthermore, from the model checking point of view, one can consider each
nominal in the input formula as a particular atomic proposition. Thus we obtain
the following result.

\begin{lemma}\label{lem:RemovingNominals}
Let $X_{\mu}$ be the $\mu$-calculus enriched with nominals and possibly with
graded modalities. Then \PDMC, \SPMC and \FSMC w.r.t. $X_{\mu}$ can be
respectively reduced in linear time to \PDMC, \SPMC and \FSMC w.r.t. $X_{\mu}$
without nominals.
\end{lemma}

\begin{proof}
In this case too, we show the proof only for \FSMC. Let $\K = \langle W, W_0,
R, L \rangle$ be a model that uses atomic propositions from $AP$ and nominals
from $Nom$, and let $\varphi$ be a formula of $X_{\mu}$. Then, we consider the
new model $\K' = \langle W, W_0, R, L \rangle$ that uses atomic propositions
from the set $AP' = AP \cup Nom$ ($\K'$ does not use nominals); moreover, let
$\varphi'$ be the formula $\varphi$ interpreted as a formula of $X_{\mu}$
without nominals on the set of atomic propositions $AP'$. Then, it is easy to
see that $\K \models \varphi$ iff $\K' \models \varphi'$.
\end{proof}

From Lemmas \ref{lem:RemovingInversePrograms} and \ref{lem:RemovingNominals}
and the fact that for propositional $\mu$--calculus \PDMC is \EXPTIME--Complete
\cite{Wal96} and \FSMC is in \UP $\cap$ \coUP \cite{Wil01}, we directly have
that hybrid $\mu$--calculus \PDMC is \EXPTIME--Complete, (full) hybrid
$\mu$--calculus \SPMC is solvable in \EXPTIME and (full) hybrid $\mu$--calculus
\FSMC is in \UP $\cap$ \coUP. Now, in \cite{Wal96} it has been showed that
$\mu$-calculus \PDMC is \EXPTIME--hard. The proof used there can be easily
adapted to handle single--push systems without incurring in any complexity
blowup. Thus, we obtain the following result.

\begin{theorem}\label{the:FullHybridMuCalculusModelChecking}
Hybrid $\mu$--calculus \PDMC is \EXPTIME--Complete, (full) hybrid
$\mu$--calculus \SPMC is \EXPTIME--Complete and (full) hybrid $\mu$--calculus
\FSMC is in \UP $\cap$ \coUP.
\end{theorem}

Finally, from Lemmas \ref{lem:RemovingInversePrograms} and
\ref{lem:RemovingNominals} we have that hybrid graded $\mu$--calculus \PDMC can
be reduced in linear time to graded $\mu$--calculus \PDMC, Fully enriched
$\mu$--calculus \SPMC can be reduced in linear time to graded $\mu$--calculus
\SPMC (note that \SPMC is a special case of \PDMC) and Fully enriched
$\mu$-calculus \FSMC can be reduced in linear time to graded $\mu$-calculus
\FSMC. Since model checking is a special case of module checking, from Theorems
\ref{the:HybridGradedFiniteStatesModuleChecking} and
\ref{the:HybridGradedPushdownModuleChecking} we have the following result.

\begin{theorem}\label{the:ModelCheckingForFullyEnrichedMuCalculus}
\PDMC is solvable in \TWOEXPTIME for (hybrid) graded $\mu$--calculus, \SPMC is
solvable in \TWOEXPTIME for Fully enriched $\mu$-calculus and \FSMC is solvable
in \EXPTIME for Fully enriched $\mu$-calculus.
\end{theorem}

\bibliographystyle{amsalpha}
\bibliography{mucalculusmodule}

\providecommand{\bysame}{\leavevmode\hbox to3em{\hrulefill}\thinspace}
\providecommand{\MR}{\relax\ifhmode\unskip\space\fi MR }
\providecommand{\MRhref}[2]{%
  \href{http://www.ams.org/mathscinet-getitem?mr=#1}{#2}
}
\providecommand{\href}[2]{#2}
\begin{thebibliography}{BLMV08}

\bibitem[AMV07]{AMV07}
A.~Aminof, A.~Murano, and M.Y. Vardi, \emph{Pushdown module checking with
  imperfect information}, CONCUR '07, LNCS, vol. 4703, Springer-{V}erlag, 2007,
  pp.~461--476.

\bibitem[Ber66]{Ber66}
R.~Berger, \emph{The undecidability of the domino problem}, Mem. AMS
  \textbf{66} (1966), 1--72.

\bibitem[BLMV06]{BLMV06}
P.A. Bonatti, C.~Lutz, A.~Murano, and M.Y. Vardi, \emph{The complexity of
  enriched $\mu$-calculi}, ICALP '06, LNCS, vol. 4052, 2006, pp.~540--551.

\bibitem[BLMV08]{BLMV08}
\bysame, \emph{The complexity of enriched $\mu$-calculi}, To appear in Logical
  Methods in Computer Science (2008), 1--27,
  http://www.na.infn.it/$\sim$murano/pubblicazioni/journal-version-enriched.pd%
f.

\bibitem[BMP05]{BMP05}
L.~Bozzelli, A.~Murano, and A.~Peron, \emph{Pushdown module checking}, LPAR,
  2005, pp.~504--518.

\bibitem[BP04]{BP04}
P.A. Bonatti and A.~Peron, \emph{On the undecidability of logics with converse,
  nominals, recursion and counting}, Artificial Intelligence \textbf{158 : 1}
  (2004), 75--96.

\bibitem[BS06]{BS06}
J.~Bradfield and C.~Stirling, \emph{Modal $\mu$-calculi}, Handbook of Modal
  Logic (Blackburn, Wolter, and van Benthem, eds.), Elsevier, 2006,
  pp.~722--756.

\bibitem[CE81]{CE81}
E.M. Clarke and E.A. Emerson, \emph{Design and synthesis of synchronization
  skeletons using branching time temporal logic}, Proc. of Work. on Logic of
  Programs, LNCS, vol. 131, 1981, pp.~52--71.

\bibitem[CGP99]{CGP99}
E.M. Clarke, O.~Grumberg, and D.A. Peled, \emph{Model checking}, MIT Press,
  Cambridge, MA, USA, 1999.

\bibitem[FM07]{FM07}
A.~Ferrante and A.~Murano, \emph{Enriched $\mu$--calculus module checking},
  FOSSACS'07, LNCS, vol. 4423, 2007, p.~183197.

\bibitem[FMP07]{FMP07}
A.~Ferrante, A.~Murano, and M.~Parente, \emph{Enriched $\mu$--calculus pushdown
  module checking}, LPAR'07, LNAI, vol. 4790, 2007, pp.~438--453.

\bibitem[Hoa85]{Hoa85}
C.A.R. Hoare, \emph{Communicating sequential processes}, 1985.

\bibitem[HP85]{HP85}
D.~Harel and A.~Pnueli, \emph{On the development of reactive systems}, Logics
  and Models of Concurrent Systems, {NATO} Advanced Summer Institutes, vol.
  F--13, Springer--Verlag, 1985, pp.~477--498.

\bibitem[Jur98]{Jur98}
Marcin Jurdzi{\'n}ski, \emph{Deciding the winner in parity games is in
  {UP}~$\cap$~co-{UP}}, Information Processing Letters \textbf{68} (1998),
  no.~3, 119--124.

\bibitem[Koz83]{Koz83}
D.~Kozen, \emph{Results on the propositional mu--calculus.}, Theoretical
  Computer Science \textbf{27} (1983), 333--354.

\bibitem[KPV02]{KPV02}
O.~Kupferman, N.~Piterman, and M.Y. Vardi, \emph{Pushdown specifications},
  LPAR, 2002, pp.~262--277.

\bibitem[KSV02]{KSV02}
O.~Kupferman, U.~Sattler, and M.Y. Vardi, \emph{The complexity of the graded
  $\mu$--calculus}, CADE '02, LNAI, vol. 2392, 2002, pp.~423--437.

\bibitem[KV97]{KV97}
O.~Kupferman and M.Y. Vardi, \emph{Module checking revisited}, CAV '96, LNCS,
  vol. 1254, Springer-{V}erlag, 1997, pp.~36--47.

\bibitem[KVW00]{KVW00}
O.~Kupferman, M.Y. Vardi, and P.~Wolper, \emph{An automata--theoretic approach
  to branching--time model checking}, Journal of ACM \textbf{47} (2000), no.~2,
  312--360.

\bibitem[KVW01]{KVW01}
\bysame, \emph{Module checking}, Information \& Computation \textbf{164}
  (2001), 322--344.

\bibitem[QS81]{QS81}
J.P. Queille and J.~Sifakis, \emph{Specification and verification of concurrent
  systems in cesar}, $5^{th}$ Symp. on Programming, LNCS, vol. 137, 1981,
  pp.~337--351.

\bibitem[SV01]{SV01}
U.~Sattler and M.Y. Vardi, \emph{The hybrid mu--calculus}, IJCAR '01, LNAI,
  vol. 2083, 2001, pp.~76--91.

\bibitem[Var98]{Var98}
M.Y. Vardi, \emph{Reasoning about the past with two--way automata}, ICALP '98,
  LNCS, vol. 1443, 1998, pp.~628--641.

\bibitem[Wal96]{Wal96}
I.~Walukiewicz, \emph{Pushdown processes: {G}ames and {M}odel {C}hecking}, CAV
  '96, LNCS, vol. 1102, Springer--{V}erlag, 1996, pp.~62--74.

\bibitem[Wal00]{Wal00}
\bysame, \emph{Model checking ctl properties of pushdown systems}, FSTTCS '00,
  LNCS, vol. 1974, Springer--Verlag, 2000, pp.~127--138.

\bibitem[Wil01]{Wil01}
T.~Wilke, \emph{Alternating tree automata, parity games, and modal
  $\mu$--calculus}, Bull. Soc. Math. Belg. \textbf{8} (2001), no.~2.

\end{thebibliography}
\end{document}